\documentclass[12pt]{article}

\usepackage{times}
\usepackage{verbatim}
\usepackage{graphicx}
\usepackage{graphics}
\usepackage{amsmath}
\usepackage{amssymb}

\title{Interaction of Phonons and Dirac Fermions on the Surface of Bi$_2$Se$_3$: A Strong Kohn Anomaly} 

\author
{Xuetao Zhu$^{1}$, L. Santos$^{2}$, R. Sankar$^3$, S. Chikara$^1$, \\C. Howard$^1$, F.C. Chou$^3$, C. Chamon$^1$, M. El-Batanouny$^{1\ast}$\\
\\
\normalsize{$^{1}$Department of Physics, Boston University, Boston, MA 02215, USA}\\
\normalsize{$^2$Department of Physics, Harvard University, Cambridge, MA 02138, USA}\\
\normalsize{$^{3}$Center of Condensed Matter Sciences, National Taiwan University, Taipei 10617, Taiwan}\\
\\
\normalsize{$^\ast$To whom correspondence should be addressed; E-mail:  elbat@bu.edu.}}

\date{}

\begin{document}
\baselineskip24pt

\maketitle

\begin{quote}
 {\bf

We report the first measurements of phonon dispersion curves on the (001) surface of the strong three-dimensional topological insulator Bi$_2$Se$_3$. The surface phonon measurements were carried out with the aid of coherent helium beam surface scattering techniques. The results reveal a prominent signature of the exotic metallic Dirac fermion quasi-particles, including a strong Kohn anomaly. The signature is manifest in a low energy isotropic convex dispersive surface phonon branch with a frequency maximum of 1.8 THz, and having a V-shaped minimum at approximately $\mathbf{2k_F}$ that defines the Kohn anomaly. Theoretical analysis attributes this dispersive profile to the renormalization of the surface phonon excitations by the surface Dirac fermions. The contribution of the Dirac fermions to this renormalization is derived in terms of a Coulomb-type perturbation model.}
\end{quote}

\newpage
The recent discovery of the new class of materials coined topological insulators (TIs) \cite{Hasan,Qi,Moore1,Fu1,Hasan2} has attracted much interest and excitement.  Strong TIs have exotic metallic surface states protected by time-reversal invariance (TRI). The presence of the metallic boundary is dictated by the recent ${\bf Z}_2$ classification of TRI insulators into an {\it ordinary} insulator class (${\bf Z}_2$-even), including the vacuum, and a {\it topological} insulator class (${\bf Z}_2$-odd). Members of each class can be adiabatically converted to each other, but not into members of the other class \cite{Kane,Moore2,Fu3}. Ordinary and topological phases are separated by a topological phase transition, where the bulk gap has to vanish at the transition point. Thus, the ensuing scenario for a strong TI depicts a topological bulk (${\bf Z}_2$-odd), embedded in vacuum (${\bf Z}_2$-even), a case which demands that the boundary (surface) be gapless. This gives rise to surface states involving massless Dirac fermion quasi-particles (DFQs) having an odd number of Dirac cones in the surface Brillouin zone (SBZ). Moreover, the strong spin-orbit coupling leads to a definite helicity whereby the spin is locked normal to the wavevector of the electronic state. A fundamental constraining feature of such spin-textured surface states is their robustness against spin-independent scattering, an attribute that protects them from backscattering and localization \cite{Roushan,Zhang3}. 

In this paper we address, experimentally and theoretically, a manifestation of the exotic surface DFQs that has received little attention: their response to surface phonon excitations, namely surface electron-phonon scattering and the ensuing phonon energy renormalization. In this work, we employ elastic and inelastic helium atom surface scattering (HASS) techniques to determine the surface structure and obtain the surface phonon dispersion curves, respectively. It is well established that He atoms are scattered by the oscillations of the surface electron density about 2-3 \AA$\ $ away from the first atomic layer, and thus HASS is very sensitive to phonon-induced surface charge density (SCD) oscillations, even those induced by subsurface second-layer displacements \cite{Chis,Benedek}. Consequently, HASS intensities carry direct information on the SCD oscillations associated with surface phonons and ultimately on the surface electron-phonon (e-p) interaction. HASS, therefore, is an ideal technique to investigate the collective Dirac electron states response to surface lattice excitations. 

Because the family of strong TIs Bi$_2$X$_3$ (X=Se, Te) was found to have a single Dirac cone with a Dirac point at the SBZ center \cite{Chen,Noh,Zhang1,Zhang2,Hsieh3},  it has been the focus of ongoing intense experimental and theoretical studies. Angle- and spin-resolved photoemission spectroscopy measurements provided strong evidence of the existence of the linearly dispersive Dirac cone and the spin texture. Moreover, scanning-tunneling microscopy confirmed the absence of backscattering, namely, scattering between states of opposite momentum and opposite spin \cite{Roushan,Park2}. We selected to study Bi$_2$Se$_3$ because of its simplicity, its relatively wide gap of 300 meV, and the rich trove of information available about its DFQ surface states \cite{Xia,Analytis,Park,Kuroda,Yazyev}. 

\begin{figure}[h!]
\begin{center}
\includegraphics[width=0.45\textwidth]{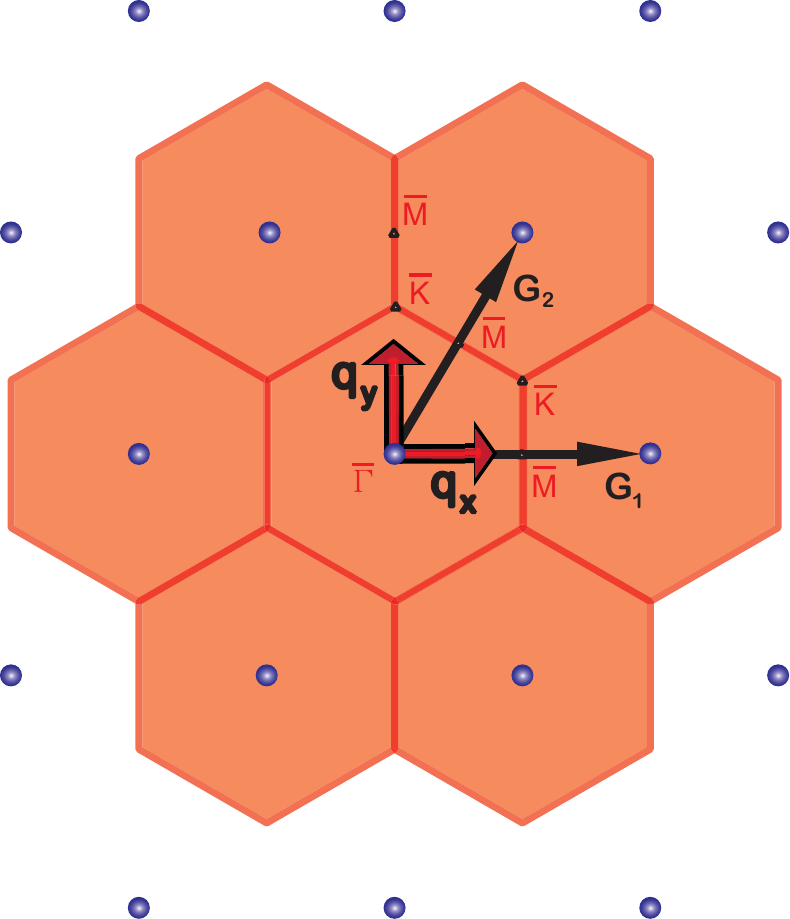}
\caption{\label{SBZ} {\small The extended surface Brillouin zone, showing  high-symmetry points $\overline{\Gamma},\ \overline{\text{M}}$ and $\overline{\text{K}}$. The $\overline{\Gamma}$-$\overline{\text{M}}$ direction lies along ${\bf q}_x$, while the $\overline{\Gamma}$-$\overline{\text{K}}$-$\overline{\text{M}}$ line is traced along ${\bf q}_y$.  The ${\bf G}_i$s are surface reciprocal lattice vectors.}}
\end{center}
\end{figure}

The (001) Bi$_2$Se$_3$ crystal surface belongs to the $p6mm$ two-dimensional space group. The corresponding SBZ is shown in figure \ref{SBZ}; with high-symmetry points $\overline{\Gamma},\ \overline{\text{M}}$ and $\overline{\text{K}}$. Extensive elastic and inelastic He scattering measurements of the Bi$_2$Se$_3$(001) surface were carried out \cite{Support}. The processed inelastic data is presented in figure \ref{spd} along the $\overline{\Gamma}$-$\overline{\text{M}}$, $\overline{\Gamma}$-$\overline{\text{K}}$ and $\overline{\text{K}}$-$\overline{\text{M}}$ directions as orange dots together with their respective error bars.

In order to interpret and fit the experimental inelastic data, we carried out phenomenological surface lattice dynamical calculations, based on the pseudo-charge model (PCM) \cite{Jayanthi,Kaden,Support} and applied to slab geometries containing 30 quintuples. The physical role of the surface DFQ states in the phonon energy renormalization was subsequently studied with the aid of a microscopic model based on Coulomb-type perturbations. The model was incorporated in calculations of density-density correlations that take into account the helicity and linear dispersion of the surface DFQ states.

\subsubsection*{Bi$_2$Se$_3$(001) surface phonon dispersions: Experimental and PCM calculations results}
The slab calculations were preceded by a bulk calculation \cite{Support} based on PCM and fitted to available Raman and IR data \cite{Richter,Landolt}. To obtain a best-fit to the measured surface phonon dispersion curves the following changes to bulk parameter values were made: The surface Se-Bi force constant was reduced by 25\% from its bulk value; this is a reasonable change since the non-metallic bonding in the two topmost layers is effectively reduced to allow for the emergence of the metallic electrons. A new planar force constant involving surface Se-Se ions was introduced. To account for the metallic deformability of the surface DFQs we reduced the magnitude of the pseudocharge-ion coupling constant $T_S$ in the topmost pyramid involving Se and Bi from its bulk value $T_B^1$ by about $\frac{\Delta T_S}{T_B^1}=\frac{T_B^1-T_S}{T_B^1}\simeq13\%$ \cite{Support}. Physically, $\Delta T_S$ accounts for the extra screening provided by the DFQ surface states, which is proportional to the corresponding Fermi surface density of states, ${\cal D}_F\propto E_F\propto k_F$. Hence $\Delta T_S\propto k_F$. 

\begin{figure}[h!]
\begin{center}
\includegraphics[width=0.8\textwidth]{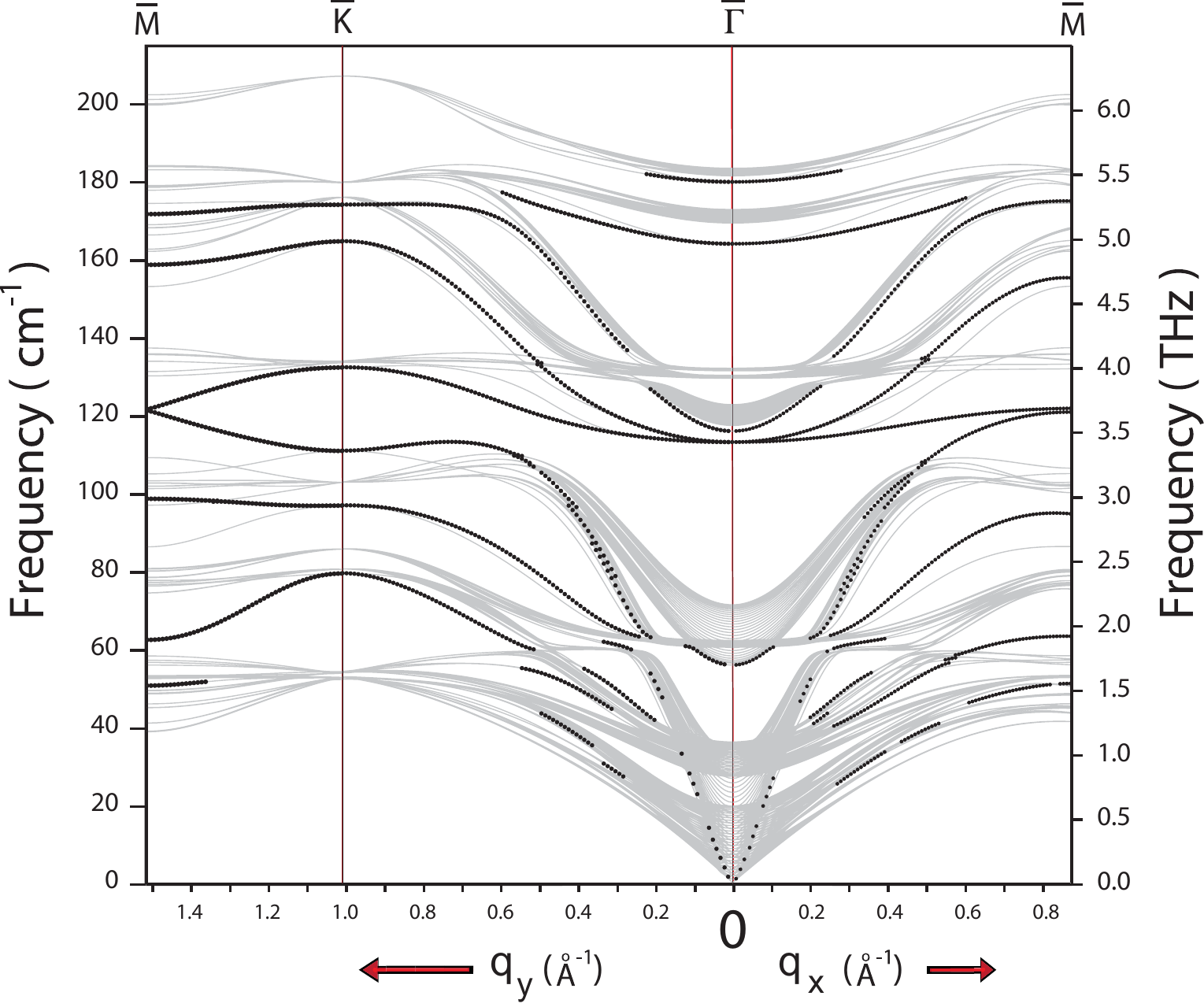}
\caption{\label{spd2} {\small Surface phonon dispersion curves (black dots) along the $\overline{\Gamma}$-$\overline{\text{M}}$ and  $\overline{\Gamma}$-$\overline{\text{K}}$ directions, for\hfill\break $T_S=T_B^1=8.07$ N/m bulk value. The gray background represents the projection of the
bulk bands on the SBZ.}}
\end{center}
\end{figure}

However, to underscore the significance of the experimental results and their interpretation in terms of PCM fitting, we start by presenting in figure \ref{spd2} the surface phonon dispersions for a model employing the bulk value of the ion-pseudo-charge coupling parameter at the surface, namely $T_S=T_B^1=8.07$ N/m \cite{Support}. The calculated surface phonon dispersion curves are presented in figure \ref{spd2} as black dots, and the gray background represents the projection of the
bulk bands onto the SBZ. The important feature to be noted is the presence of a surface Rayleigh branch extending from $\omega=0$ to $\omega\simeq3.7$ THz.

\begin{figure}[h!]
\begin{center}
\includegraphics[width=0.75\textwidth]{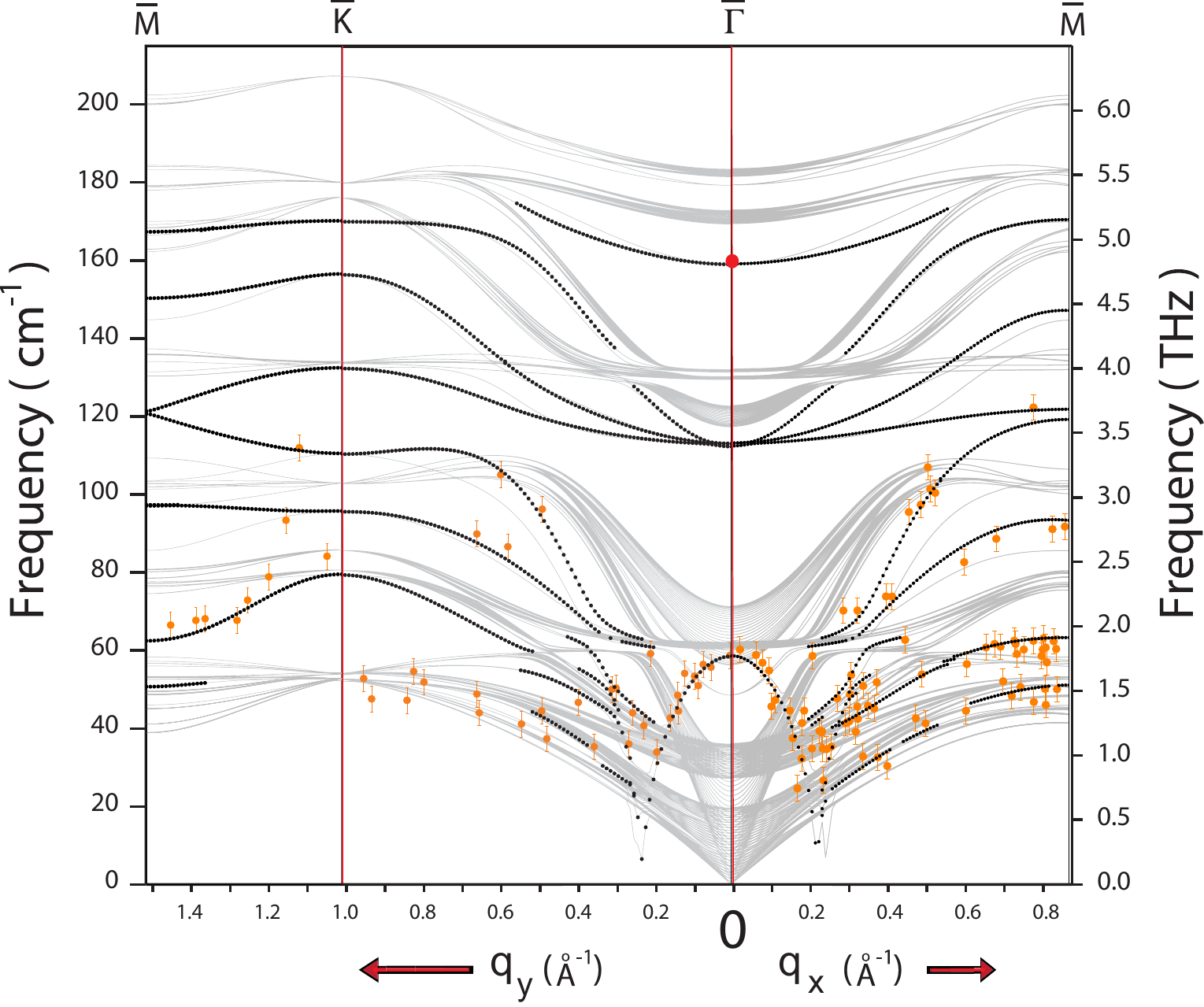}
\caption{\label{spd} {\small Surface phonon dispersion curves along the $\overline{\Gamma}$-$\overline{\text{M}}$ and  $\overline{\Gamma}$-$\overline{\text{K}}$ directions: Experimental data appear as orange dots with error
bars reflecting instrument resolution, while the calculated dispersions curves, using PCM with $T_S=7.05$ N/m, are represented by black dots. The gray background represents the projection of the
bulk bands on the SBZ. The red dot on $\overline{\Gamma}$ at 160
cm$^{-1}$ represents an experimental surface Raman mode reported in
Ref.{\cite{Zhao}}. }}
\end{center}
\end{figure}

By contrast, the calculated surface phonon dispersion curves that provide a best-fit for the experimental data, correspond to a reduced value of $T_S$ of 7.05 N/m. The calculated surface phonon dispersion curves are shown in figure \ref{spd} as black dots, superimposed on the experimental data which is displayed as solid orange dots with error bars. The single red dot at the $\overline{\Gamma}$-point, at about 5 THz, is a Raman surface frequency reported in \cite{Zhao}.

We shall focus our attention on two important observations regarding figure \ref{spd}. First, we should discern the absence of a surface Rayleigh branch in both experimental and calculated dispersion curves. Second, we observe the emergence of a ``prominent" isotropic parabolic dispersion branch centered at the $\overline{\Gamma}$-point with a frequency of 1.8 THz (7.4 meV); the PCM calculations confirm that at $\overline{\Gamma}$ it has a  vertical shear (z) polarization, with $u^z_{Se}/u^z_{Bi}=2.8$, where $u^z$ is the respective ionic displacement. This branch terminates in a V-shaped minimum located at $q\simeq 0.2$ \AA$^{-1}$, a value that roughly corresponds to $2k_F$ of the DFQs, and thus signals the manifestation of a strong Kohn anomaly \cite{Kohn}. The isotropy of this branch and its apparent termination at $2k_F$ can be explained by a scenario involving the DFQ surface states, in particular their isotropic Fermi surface. In this scenario, the V-shaped feature marks the boundary between an operative DFQ screening for $q<2k_F$, and its suppression above this value, which is a typical signature of a Kohn anomaly. Lattice dynamics calculations reveal some bulk penetration of vertical shear modes for $q>2k_F$ reflecting a diminished role of DFQ screening and more compatibility with the insulating bulk. We shall establish below the intimate link between the dispersive character of this branch and the surface DFQ states response to ionic displacements. 
 
The polarization and other properties of the remaining surface phonon dispersive branches are discussed in \cite{Support}.
\subsubsection*{Perturbative Coulomb interactions results}
We now briefly describe the microscopic model used to study the contributions of the surface Dirac electronic states that define the dispersive character of the prominent phonon branch. Our experimental results demonstrate that this surface phonon branch is ``optical" in nature, and thus should be derived from a Coulomb-type perturbative mechanism. Consequently, we construct a perturbative Hamiltonian describing the interaction of the surface Dirac electrons with the electric field resulting from ionic displacements, namely, a linear coupling of the lattice ionic displacement to the DFQ density of  the form 
\begin{equation}
\label{eq:electron-phonon interaction}
{\cal H}_{\rm{el-ph}}
=
\int\,\mathrm{d}^{2}{\bf r}\,\rho_{\rm{el}}({\bf r})\,\sum_{j=1}^{N}\,\boldsymbol{\lambda}(\,{\bf r}-{\bf R}^{(0)}_{j}\,)\,\boldsymbol{\cdot}\mathbf{u}_j\;,
\end{equation}
where  $\rho_{\rm el}({\bf r})$ is the 2-dimensional electron density, and $\mathbf{u}_j$ is the displacement of the $j^{\rm th}$ surface ion whose equilibrium position is ${\bf R}^{(0)}_{j}$.  $\boldsymbol{\lambda}$ is a vector accounting for the position-dependent coupling of the charge density to ionic displacements. As shown in \cite{Support}, after second quantization, with $b_{{\bf q},\gamma}$ and $c_{{\bf k},\,\alpha}$ being the annihilation operators for a surface phonon mode $({\bf q},\gamma)$ with polarization index $\gamma$ and an electron with momentum ${\bf k}$ and spin $\alpha$, respectively, the Hamiltonian takes the form
\begin{equation}
{\cal H}_{\rm{el-ph}}
=
\frac{1}{\sqrt{\mathcal{A}}}\,\sum_{\alpha = \uparrow,\downarrow}\,\sum_{{\bf k}\atop{\bf q},\gamma}\;{\rm g}_{{\bf q},\gamma}\,
c^{\dagger}_{{\bf k}+{\bf q},\,\alpha}\,c_{{\bf k},\,\alpha}\;\hat{A}_{{\bf q},\gamma}\;,
\end{equation}
where $\hat{A}_{{\bf q},\gamma}\,\equiv\,(b_{{\bf q},\gamma}+b^{\dagger}_{{\bf q},\gamma})$. The electron-phonon coupling constant is 
\begin{equation}
\label{eq:electron-phonon coupling constant general}
{\rm g}_{{\bf q},\gamma} = \sqrt{\frac{N\hbar}{2 M \mathcal{A}\, \omega^{(0)}_{{\bf q},\gamma}}}\;
\boldsymbol{\lambda}_{\mathbf q} \boldsymbol{\cdot}\hat{\bf e}_{\gamma}(\mathbf q)	
\equiv
 \sqrt{\frac{N\hbar}{2 M \mathcal{A}\, \omega^{(0)}_{{\bf q},\gamma}}}\;
\lambda_{{\mathbf q},\gamma}\;.
\end{equation} $\mathcal{A}$ is the surface area, $N$ the number of primitive cells in ${\cal A}$, $M$ the ionic mass. $\omega^{(0)}_{{\bf q},\gamma}$ is the bare phonon frequency of mode $({\bf q},\gamma)$ and $\hat{\bf e}_{\gamma}(\mathbf q)$ the corresponding polarization vector. $\boldsymbol{\lambda}_{\mathbf q}$ denotes the Fourier transform of $\boldsymbol{\lambda}({\bf r})$.

A detailed derivation given in \cite{Support} leads to the Dyson equation
\begin{equation}
\label{eq:self-consistent renormalized frequency}
\Bigl(\hbar\omega_{{\bf q},\gamma}\Bigr)^2
=
\Bigl(\hbar\omega^{(0)}_{{\bf q},\gamma}\Bigr)^2 
+\frac{\hbar^2}{M\mathfrak{A}}\,|\lambda_{{\bf q}}|^{2}\,
\frac{\Pi({\bf q}, \omega_{{\bf q},\gamma})}{\varepsilon({\bf q}, \omega_{{\bf q},\gamma})}\;,	
\end{equation} where $\omega_{{\bf q},\gamma}$ is the renormalized surface phonon frequency, and $\mathfrak{A}$ is the surface primitive cell area. $\Pi$ and $\varepsilon$ are the polarization and dielectric functions in the random phase approximation, respectively, expressed in the helicity basis; the former can be decomposed into two contributions: one due to intra-band and the other due to inter-band excitations as shown in figure \ref{DC}:
\begin{figure} 
\begin{center}
\includegraphics[width=0.5\textwidth]{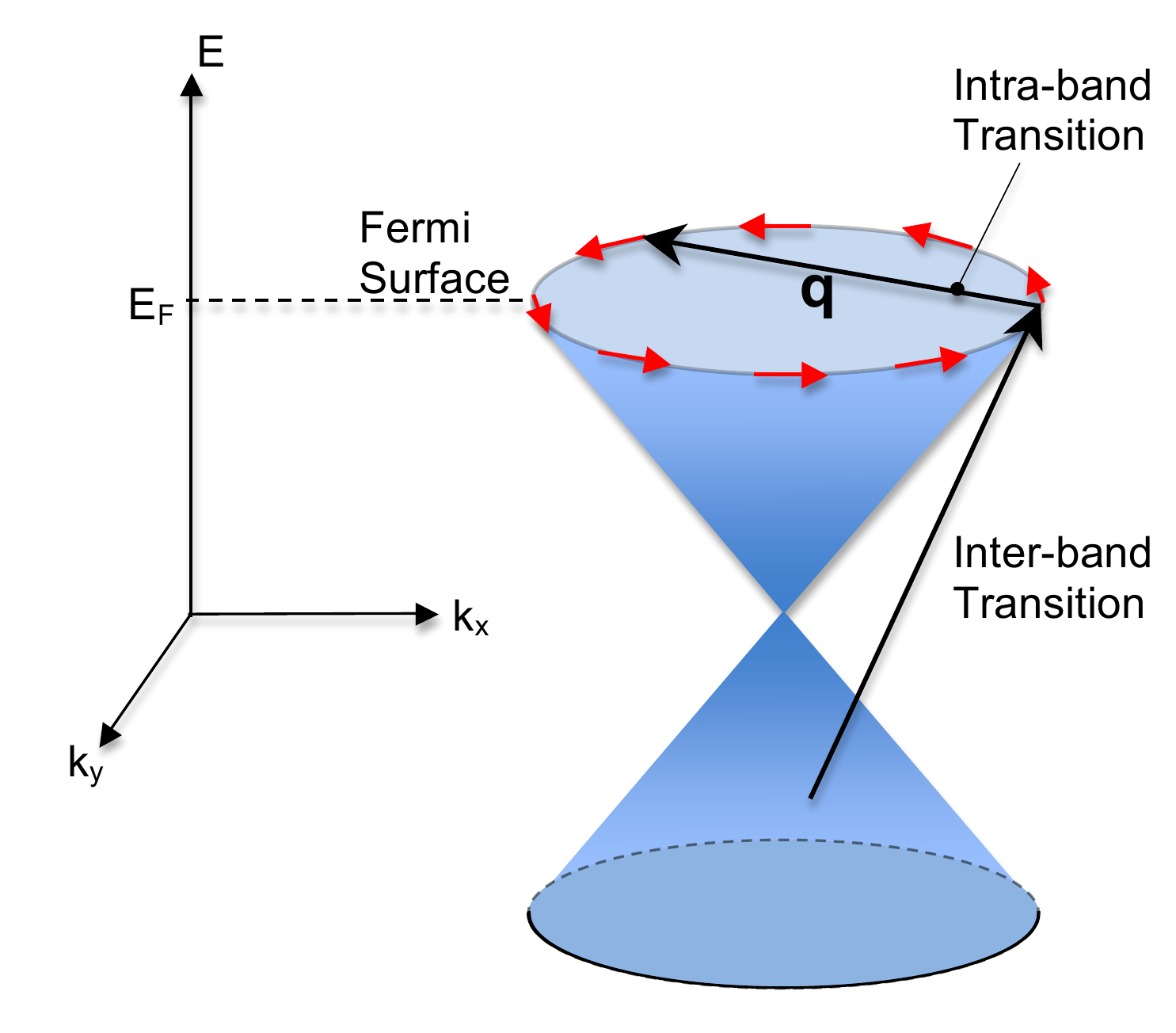}
\caption{\label{DC} {\small Intra- and Inter-band transitions of DFQs that contribute to the renormalization of the prominent surface phonon branch. ${\bf q}$ is the phonon wave vector.}}
\end{center}
\end{figure}

\begin{equation}
\Pi({\bf q},\omega)
=
\Pi^{\text{intra}}({\bf q},\omega)
+
\Pi^{\text{inter}}({\bf q},\omega)\;.	
\end{equation}
Explicit expressions for $\Pi^{\text{intra}}$ and $\Pi^{\text{inter}}$ are given in \cite{Support}.

In order to solve (\ref{eq:self-consistent renormalized frequency}) self consistently,
we need to specify the momentum dependence of the coupling function $\boldsymbol{\lambda}_{\mathbf{q},\gamma}$.
We argue that in the range of $\left|{\bf q}\right|\le 2k_F$, the ionic screened potential $V({\bf q})$  has very weak dependence on $q$, since $2k_F < q_{\rm TF}=\frac{e^2\,k_{F}^2}{4\pi\varepsilon_{0}\,E_{F}}=0.5$ \AA$^{-1}$ the Thomas-Fermi wave vector. Moreover, invoking the sagittal-plane symmetry classification of the surface phonon modes, we specify the general polarization vector of the even parity mode as
\begin{equation}
\hat{\bf e}_{\gamma}(\mathbf{q})
\Rightarrow\,
\hat{\bf e}_{\bot}(\mathbf{q}) \sim \hat{\mathbf{z}},
\quad
\hat{\bf e}_{\|}(\mathbf{q}) \sim \hat{\mathbf{q}}\;.
\end{equation}
$\hat{\bf e}_{\bot}(\mathbf{q})$ accounts for the vertical shear component (ionic oscillations
 in a direction normal to the surface plane) while
$\hat{\bf e}_{\|}(\mathbf{q})$ is the longitudinal component.
The corresponding coupling constant then reads
\begin{align}
\lambda_{\mathbf{q},\gamma}
&=
\boldsymbol{\lambda}_{\mathbf{q}}\boldsymbol{\cdot}\hat{\bf e}_{\gamma}(\mathbf{q}\,)\,
=\,
\boldsymbol{\lambda}_{\mathbf{q}}\boldsymbol{\cdot}\hat{\bf e}_{\bot}(\mathbf{q}\,)
+
\boldsymbol{\lambda}_{\mathbf{q}}\boldsymbol{\cdot}\hat{\bf e}_{\|}(\mathbf{q}\,)\notag
\\
&
\equiv
\lambda_{\bot}(\mathbf{q}) + \lambda_{\|}(\mathbf{q})\;.
\end{align}
In view of the near constancy of $V({\bf q})$ for $q\le 2k_F$, and the fact that the electron-phonon coupling involves the gradient of the screened potential, we write
\begin{equation}
\lambda_{\mathbf{q}}
=
\lambda^{(0)}_{\bot}
+
\frac{|\mathbf{q}|}{q_{0}}\lambda^{(0)}_{\|}	
\end{equation}
where $q_0$ will be appropriately chosen as $2k_F$. Notice that at $\overline{\Gamma}$ (${\bf q}={\bf 0}$) the polarization is pure vertical sheer, in agreement with the surface symmetry and the PCM.

 The Dyson equation now becomes
\begin{equation}
(\hbar
\omega_{{\bf q},\gamma})^2=(\hbar\omega^{(0)}_{{\bf q},\gamma})^2+\frac{\hbar^2}{M\mathfrak{A}}\,(\lambda_{\bot})^2\left(1+\frac{|\mathbf{q}|}{k_F}\frac{\lambda_{\|}}{\lambda_{\bot}}\right)\,\frac{\Pi({\bf q},\omega_{{\bf q},\gamma})}{\varepsilon({\bf q},\omega_{{\bf q},\gamma})}
\end{equation}
to first order in $q$.

\begin{figure} 
\begin{center}
\includegraphics[width=0.5\textwidth]{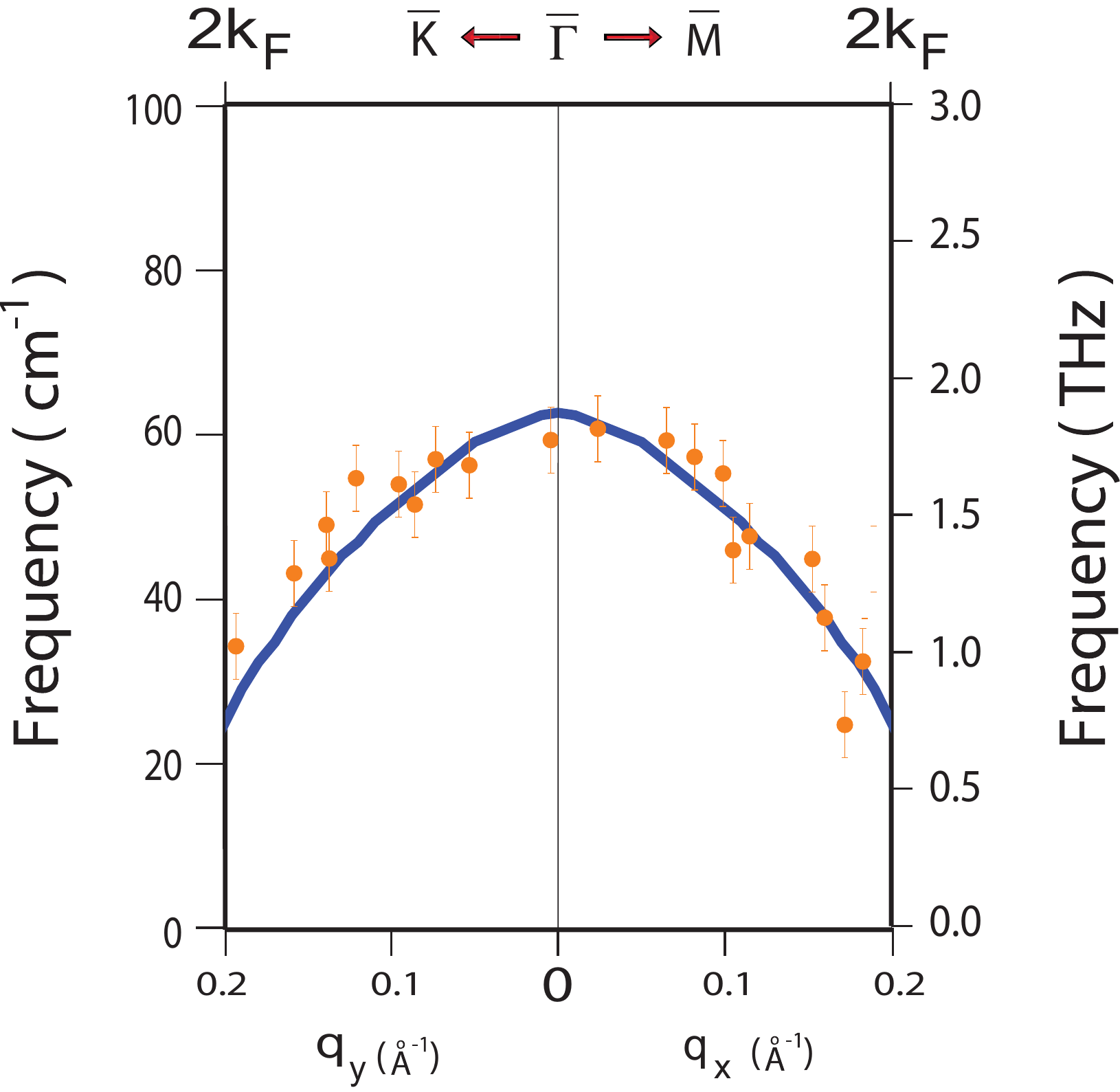}
\caption{\label{Th} {\small Renormalized surface topological phonon dispersion curve, superimposed on the corresponding experimental data.}}
\end{center}
\end{figure}

The model depends on two parameters: the bare phonon frequency and the Fermi wave vector. We identified the former with the experimental value of $\omega({\bf q}=0)=1.8$ THz, where the DFQ response vanishes. $k_F$ was derived from a sample carrier concentration of $-1.9\times10^{19}/{\rm cm}^3$, obtained from Hall measurements, which correspond to a Fermi energy of about 300 meV and a Fermi wavevector of $k_F=0.1$ \AA$^{-1}$, which is consistent with previous values reported from photoemission measurements \cite{Xia,Park,Kuroda}. The coupling parameters $\lambda^{(0)}_{\bot}$ and $\lambda^{(0)}_{\|}$ were left as fitting variables. Solutions for $\omega({\bf q})$ were obtained by carrying out the integrals for $\Pi({\bf q},\omega)$, and solving Eq.~(\ref{eq:self-consistent renormalized frequency}) iteratively. The calculated best-fit dispersion curve is shown in figure \ref{Th}, superimposed on the experimental data. The agreement is quite good.

The best fit parameters are
\begin{equation}
\frac{\hbar^2}{M\mathfrak{A}}\,(\lambda_{\bot})^2=10^7 \
\left({\rm meV}\right)^3\cdot{\textup{\AA}}^2, \qquad
\frac{\lambda_{\|}}{\lambda_{\bot}}=0.65\;.
\end{equation}
With $\frac{\hbar^2}{M\mathfrak{A}}=4\times 10^{-3}$ meV, we obtain
$$\lambda_{\bot}=50 \ eV\cdot{\textup{\AA}}\;.$$ This yields a real space value of 3.4 eV/{\textup{\AA}}, which is quite reasonable since it falls in the range of typical Coulombic interactions.

Finally, we demonstrated above that a TI actually presents a composite system consisting of an ``ultra-thin metallic film" and an underlying insulating substrate. As such, the phonons in the metallic film can hardly penetrate into the substrate bulk in order to establish an acoustic Rayleigh branch. However, the unique feature of this system is that for $q>2k_F$, as DFQ screening becomes gradually suppressed, the surface film becomes almost homogeneous with the underlying substrate, and the corresponding modes gradually penetrate into the bulk, ushering the V-shaped dispersion. We designate this unique behavior as a {\sl strong} Kohn anomaly. Moreover, we should emphasize here that theoretical modeling and analysis of the interaction between the surface phonons and the DFQs demonstrate that  the linear dispersion and isotropy of the DFQs are responsible for the profile of the prominent surface phonon branch for $q\le 2k_F$.

\section*{Methods}
\subsection*{Theoretical Calculations}
The pseudo-charge model (PCM) was incorporated into a slab-geometry program to calculate the surface phonon dispersion curves. PCM includes the electronic degrees of freedom as massless particles (adiabatic approximation) with dynamical variables associated with symmetry-adapted deformation functions. Simple elimination of these dynamical variables leads to an effective screening of the ionic motions.
\subsection*{Experimental Methods}
Bi$_2$Se$_3$ single crystals were prepared using the vertical Bridgman method.  The initial stoichiometric mixtures of Bi and Se powder of 5N purity were sealed in an evacuated conical quartz ampoule. Typical crystal samples used were about 3mm$\times$3mm$\times$2mm in size, and were cleaved in situ under ultra-high vacuum conditions, with the aid of an attached post. Inelastic HASS measurements were carried out with an angle-resolved detector equipped with a time-of-flight (TOF) system using a gating technique based on electronic excitations of helium atoms to the 2$^3$S triplet metastable state. Data processing of TOF spectra was done with the aid of {\it scan curves}, namely $\Delta E(${\small$ \Delta K)$}, where $\Delta E$ and {\small $\Delta K$} are the energy exchange and momentum transfer, respectively.

A more detailed account of the Methods is given in the {\it supplementary material}.
\newpage
\section*{Supplementary Material}

\subsection*{1. The topological insulator Bi$_2$Se$_3$}
\begin{figure}[h!]
\begin{center}
\includegraphics[width=1.0\textwidth]{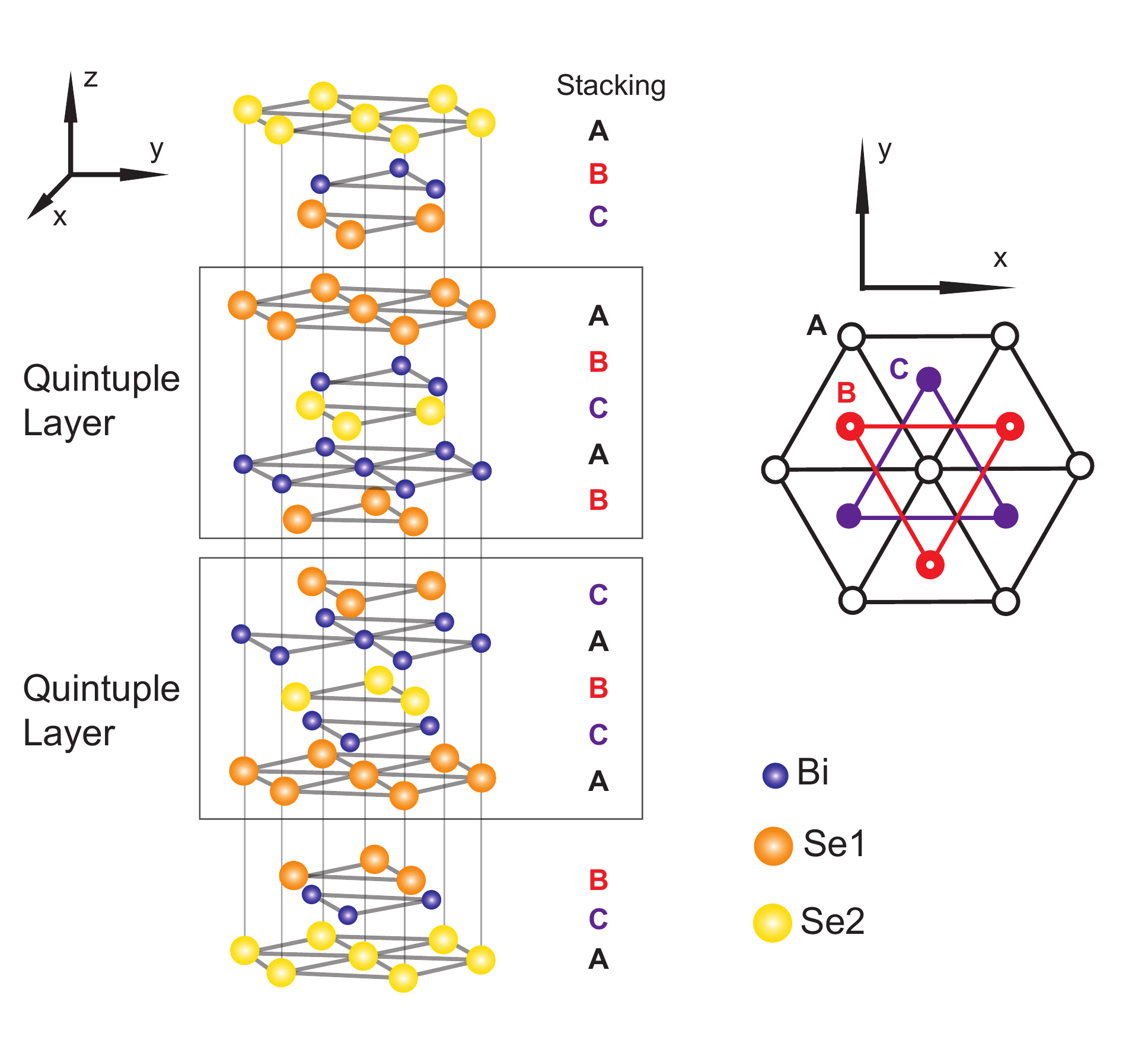}
\caption{\label{cs} {\small Crystal Structure of Bi$_2$Se$_3$. (Left) A unit cell showing the quintuple structure. (Right) A top view showing the stacking arrangement.}}
\end{center}
\end{figure}

\begin{figure}[h!]
\begin{center}
\includegraphics[width=1.0\textwidth]{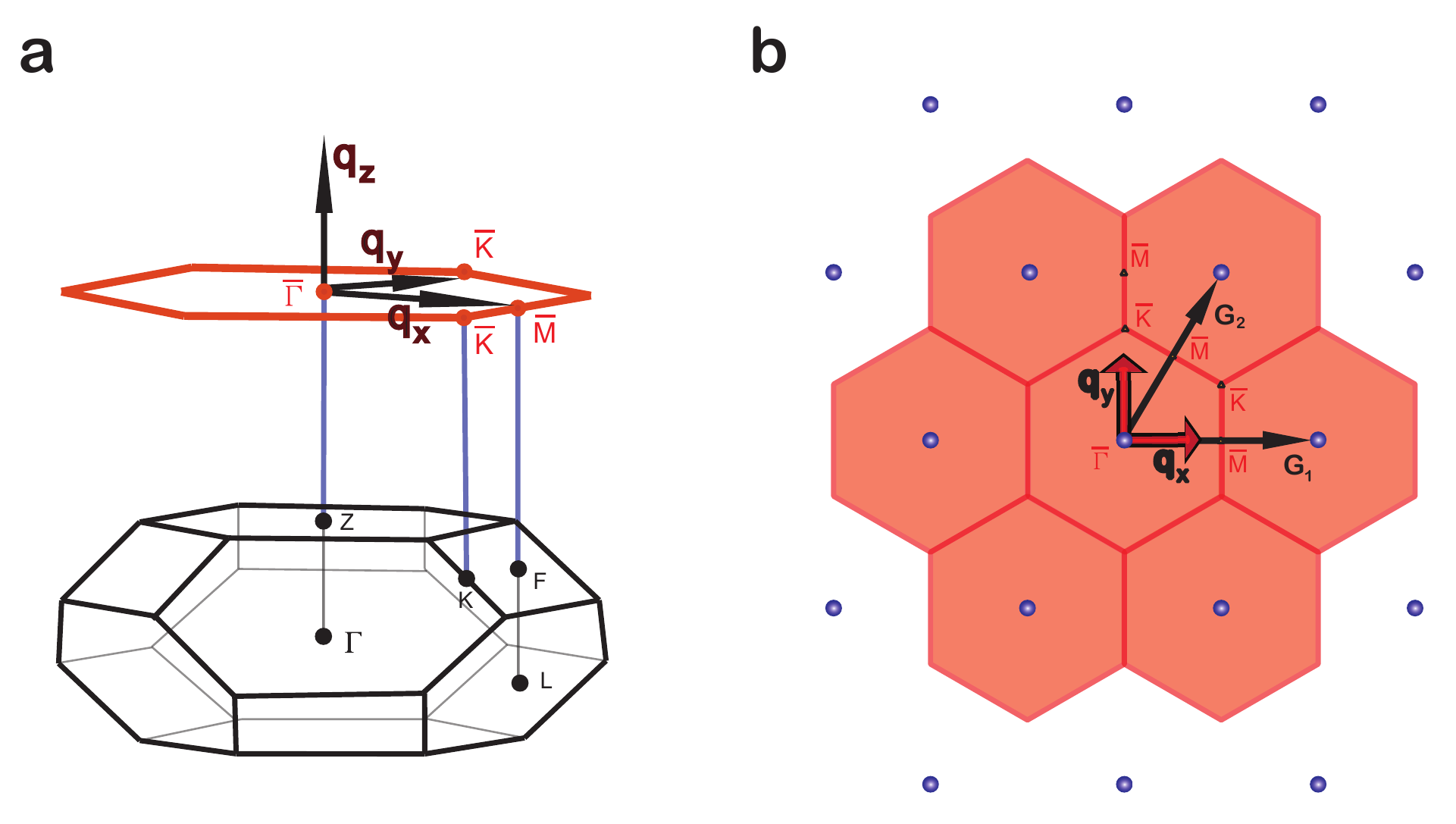}
\caption{\label{bzs} {\small Brillouin Zones (BZ). ({\bf a}) A schematic view of the bulk three-dimensional BZ of Bi$_2$Se$_3$ and the two-dimensional BZ of the projected (001) surface. ({\bf b}) The extended surface BZ with reciprocal lattice vectors ${\bf G_1}$ and ${\bf G_2}$. Experimental measurements were carried out along the two high symmetry directions, $\overline{\Gamma}$-$\overline{\text{M}}$ and  $\overline{\Gamma}$-$\overline{\text{K}}$, which are aligned along ${\bf q_x}$ and ${\bf q_y}$, respectively.}}
\end{center}
\end{figure}

Bi$_2$Se$_3$ has a rhombohedral crystal structure with space group ${\bf D}^5_\text{3d}$ (${\bf R}\bar{3}m$). The primitive cell, shown in figure \ref{cs}, contains a binary axis (with twofold rotation symmetry), a bisectrix axis (appearing in the reflection plane) and a trigonal axis (with threefold rotation symmetry). The primitive translation vectors in the rhombohedral and hexagonal basis are

{\small\begin{align}{\bf t}^R_1\, &=\,\left(-\frac{a}{2},\,-\frac{\sqrt{3}a}{6},\, \frac{c}{3}\right),\ {\bf t}^R_2\,=\,\left(\frac{a}{2},\,-\frac{\sqrt{3}a}{6},\, \frac{c}{3}\right),\ {\bf t}^R_3\, =\,\left(0,\,\frac{\sqrt{3}a}{6},\, \frac{c}{3}\right),\notag\\{\bf t}^H_1\, &=\,\left(a,\,0,\, 0\right),\ {\bf t}^H_2\,=\,\left(\frac{a}{2},\,\frac{\sqrt{3}a}{2},\, 0\right),\ {\bf t}^H_3\, =\,\left(0,\,0,\, c\right),\end{align}}
where $a$ and $c$ are lattice constants of the hexagonal cell. The corresponding reciprocal vectors are given as
{\small\begin{align}{\bf b}^R_1\,&=\,\left(-1,\,-\frac{\sqrt{3}}{6},\, \frac{a}{c}\right)\,\frac{2\pi}{a},\ {\bf b}^R_2\,=\,\left(1,\,-\frac{\sqrt{3}}{6},\, \frac{a}{c}\right)\,\frac{2\pi}{a},\ {\bf b}^R_3\,=\,\left(0,\,\frac{2\sqrt{3}}{6},\, \frac{a}{c}\right)\,\frac{2\pi}{a},\notag\\{\bf b}^H_1\,&=\,\left(1,\,\frac{1}{\sqrt{3}},\, 0\right)\,\frac{2\pi}{a},\ {\bf b}^H_2\,=\,\left(0,\,\frac{2}{\sqrt{3}},\, 0\right)\,\frac{2\pi}{a},\ {\bf b}^H_3\,=\,\left(0,\,0,\, 1\right)\,\frac{2\pi}{c}.\end{align}} The bulk Brillouin zone and surface Brillouin zone (SBZ) are shown in figure \ref{bzs}.a; while an extended version of the SBZ is displayed in figure \ref{bzs}.b. 
Bi$_2$Se$_3$ displays a layered structure consisting of five atomic layers as a basic unit (cell), ordered in the Se(1)-Bi-Se(2)-Bi-Se(1) sequence, designated a quintuple layer (QL), as shown in figure \ref{cs}. Se(2) is a center of inversion. Within each QL, the inter-layer bonding is strong and exhibits a predominant covalent character. The inter-quintuple bonding is of the van der Waals-type. 

\subsection*{2. Experimental Setup and Procedures} 

\subsubsection*{A. Crystal Preparation}   
Bi$_2$Se$_3$ single crystals were prepared using the vertical Bridgman method.  The initial stoichiometric mixtures of Bi and Se powder of 5N purity were sealed in a conical quartz ampoule and evacuated to 10 mTorr after repeated Argon gas purging. Preliminary synthesis and homogenization was carried out in a horizontal tube furnace at 650 C for 48 hours. The sealed ampoules were then passed through a vertical Bridgman furnace with a hot zone at 850 C and a thermal gradient of 2 C/cm near the solidification point 705 C.  The pulling rate was kept at 0.2 mm/hr after soaking at 850 C for 24 hours. The grown single crystals were 3 cm long and about 1.2 cm in diameter, with good optical quality. They are easy to cleave with crystal planes perpendicular to the trigonal c-axis. Hall conductivity measurements on these samples gave carrier concentration of about $-1.9\times10^{19}/{\rm cm}^3$.

 Typical crystal samples used were about 3mm$\times$3mm$\times$2mm in size, with its exposed surface parallel to the QL planes. The crystals were attached to an OFHC copper sample-holder by conductive silver epoxy. A cleaving (peeling) post was attached to the top sample surface in a similar way. The prepared sample holder was mounted on a sample manipulator equipped with XYZ motions as well as polar and azimuthal rotations. The pressure in the Ultra-High Vacuum (UHV) chamber was maintained at $10^{-10}$ torr throughout the experiment to ensure the cleanliness of the sample surface during measurement. In situ cleaving under UHV conditions was
effected by knocking off the cleaving post. All measurements were performed with the sample surface at room temperature. Immediately after cleaving, the quality of the long-range ordering on the surface was confirmed by the appearance of sharp diffraction LEED spots.
\subsubsection*{B. Helium beam monochromator}
A supersonic mono-energetic collimated helium beam, with velocity resolution better than\hfill\break 1.4 $\%$, was generated by a nozzle-skimmer
assembly and 2mm diameter collimating slits. The average beam velocity was varied by attaching the nozzle reservoir to a closed-cycle helium refrigerator, and controlling the reservoir temperature, in the range 300 K to 110 K, with the aid of a digital temperature controller (Scientific Instruments Model 9700) and a diode sensor attached to the reservoir. This allows the beam energy to vary in the range 65 meV to 21 meV. Polar rotation of the sample was used to vary the incident angle $\theta_{i}$ with respect to the surface normal, while the azimuthal rotation was employed to align the scattering plane along a high-symmetry surface crystallographic direction. 
\subsubsection*{C. Scattered beam detection technique}

\begin{figure}[h!]
\begin{center}
\includegraphics[width=0.75\textwidth]{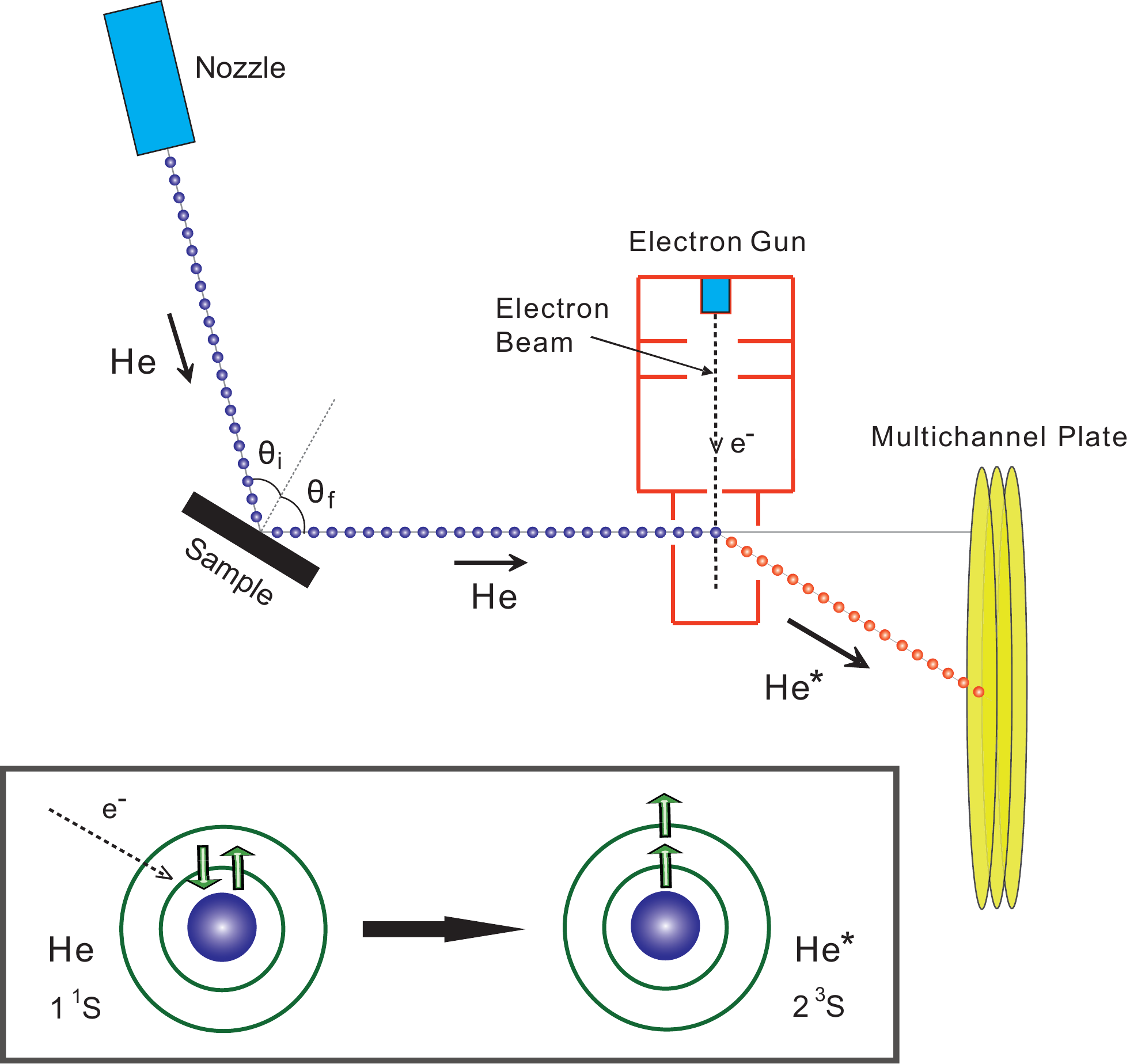}
\caption{\label{det}{\small Schematic of the He beam scattering geometry, showing the metastable He$^*$ atom detection scheme. }}
\end{center}
\end{figure}

The scattered He beam was collected by an angle-resolved detector mounted on a two-axis goniometer, which allows the scattered angle $\theta_{f}$ to be varied independently from $\theta_{i}$ \cite{Farzaneh}, and allows in- and out-of the scattering-plane measurements. As shown in figure \ref{det}, the detector is comprised of an electron gun and a multichannel plate (MCP) electron multiplier \cite{Martini}. The electron gun generates a well-collimated, monoenergetic electron beam crossing the He beam at right angles. The energy of the electron beam is tuned to excite the He atoms to their first excited metastable state 2$^3$S (He*) upon impact. Deexcitation of a He* atom at the surface of the MCP leads to the ejection of an electron (Penning ionization) which generates an electron cascade that is collected by the anode of the multiplier. By electronically pulsing the electron gun, a gate function is created for time-of-flight (TOF) measurements in the inelastic HASS mode. The details of the detection scheme are given in Ref \cite{Martini}.

 Inelastic scattering measurements were carried out for in- and out-of- the-scattering-plane geometry. By writing the He-atom wave vector as ${\bf k}=({\bf K},k_z)$, where ${\bf K}$ is the component parallel to the surface, conservation of momentum and energy for in-the-scattering-plane geometry can be expressed as

{\small\begin{align}\label{1a}
\Delta{\bf K}\,&=\,{\bf G}+{\bf Q}\,=\,k_{f}\sin\theta_{f}-k_{i}\sin\theta_{i}\\\label{2a}
\Delta E\,&=\,\hbar\omega({\bf Q})\,=\,\frac{\hbar^2k_f^2}{2M_\text{He}}-\frac{\hbar^2k_i^2}{2M_\text{He}}
\end{align}}
where subscripts $i$ and $f$ denote incident and scattered beams, respectively, and $\Delta{\bf K}$ is the momentum transfer parallel
to the surface. ${\bf G}$ is a surface reciprocal-lattice vector, ${\bf Q}$ is the surface phonon wave vector, $\hbar\omega({\bf
Q})$  is the corresponding surface phonon energy and $M_\text{He}$ is the mass of a He atom. By eliminating ${\bf k}_f$ from the above equations, one obtains the so-called {\sl scan curve} relations which are the locus of all the allowed $\Delta{\bf K}$ and $\Delta E$ as dictated by the geometry and the conservation relations,

{\small\begin{equation} \Delta E\,=\,E_i\,\left[\left(\frac{\sin\theta_i + \Delta{\bf K}/k_i}{\sin\theta_f}\right)^2-1\right]
\end{equation}}
where $E_i=\hbar^2k_i^2/2M_\text{He}$. The intersections of these scan curves with the phonon dispersion curves define the kinematically allowed inelastic events for a fixed geometric arrangement. Thus, by systematically changing $E_i,\,\theta_i$, and $\theta_f$, the entire set of dispersion curves can be constructed.
\begin{figure}[h!]
\begin{center}
\includegraphics[width=0.9\textwidth]{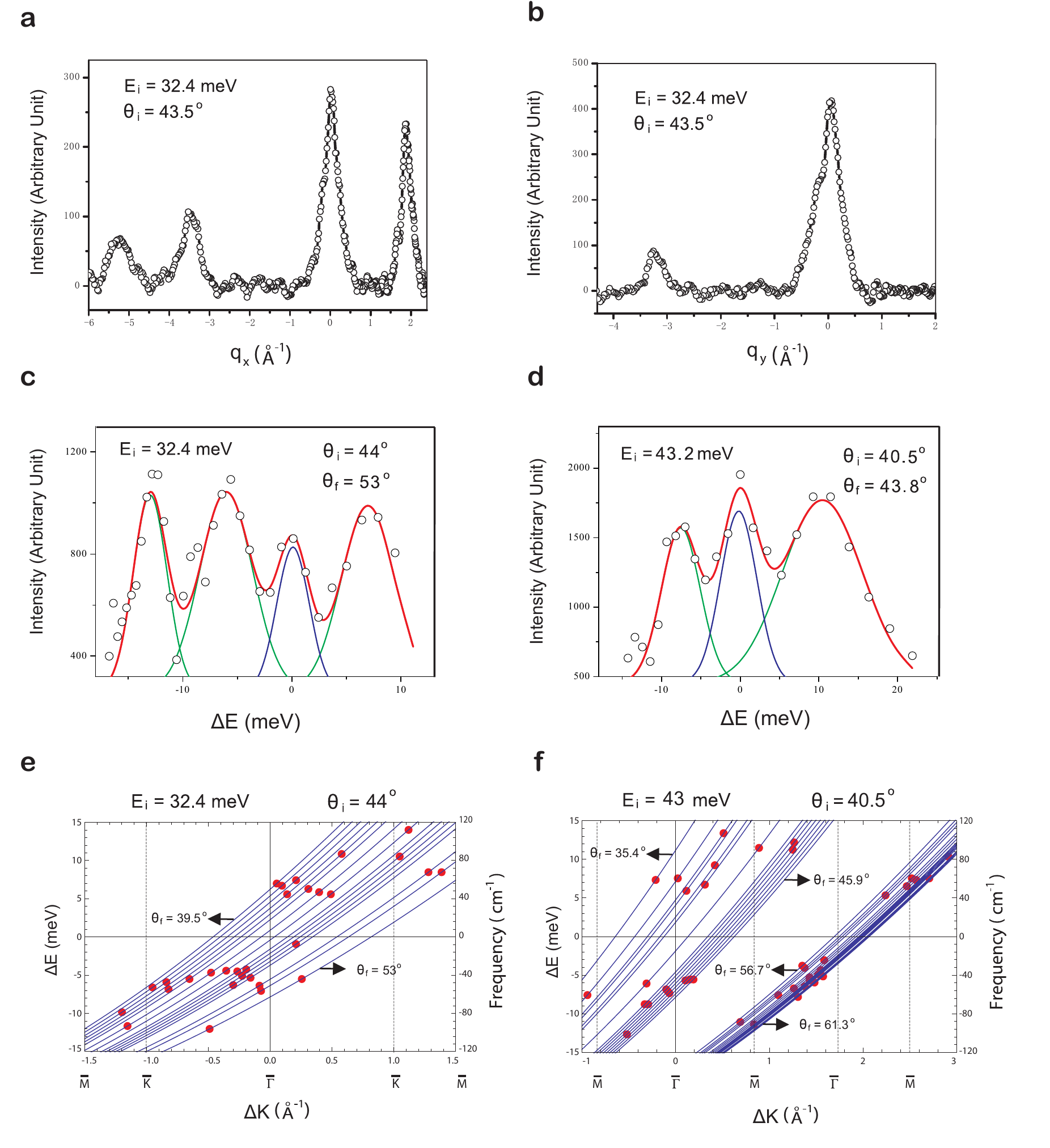}
\caption{\label{res}{\small Typical experimental results from HASS
measurements. {\bf a-b}, Elastic diffraction patterns along
$\overline{\Gamma}$-$\overline{\text{M}}$ ({\bf a}),
and  $\overline{\Gamma}$-$\overline{\text{K}}$
({\bf b}) directions.
{\bf c-d}, Typical energy loss (gain) spectra from inelastic
TOF measurements. Experimental data
(circles) are resolved into inelastic peaks (green) and diffuse
elastic peaks (blue). {\bf e-f}, Scan curves with TOF 
experimental results (red dots). Each red dot represents
a surface phonon event, and also appears as a
green peak in {\bf c-d}. }}
\end{center}
\end{figure}

\subsubsection*{D. Typical Experimental Results}
Figure \ref{res} (a) and (b) show typical diffraction patterns along the $\overline{\Gamma}$-$\overline{\text{M}}$ and $\overline{\Gamma}$-$\overline{\text{K}}$ directions, confirming the hexagonal geometry of the surface and the lattice constant $a=4.15$ \AA. Figure \ref{res} (c) and (d) show typical TOF spectra displaying  diffusive elastic as well as inelastic peaks; Gaussian fits to the displayed peaks are shown in blue for the diffuse elastic peak, and in green for the inelastic ones. Figure \ref{res} (e) and (f) display typical scan curves with experimental inelastic events shown as red dots. 
\subsection*{3. Pseudo-charge Phonon Model \cite{Jayanthi,Kaden,Benedek}}
In order to identify the character of the measured phonon dispersion curves in terms of irreducible representations (Irreps) and polarization vectors, we use an empirical lattice-dynamical model, the pseudo-charge model, that includes both Born-von K\'{a}rm\'{a}n type direct ion-ion force constants, as well as indirect ion-ion adiabatic coupling through the mediating electrons. This model has been particularly successful in reproducing both the surface dispersion curves and the HASS
scattering amplitudes, derived from calculated surface charge deformations. 

The model accounts explicitly for the electronic degrees of freedom in the following manner: The electron density $n_\ell$ within each primitive cell $\ell$ is expanded in terms of symmetry-adapted (SA) multipole components around selected Wyckoff symmetry points ${\bf R}_{\ell,j}$, so that the corresponding {\sl pseudo-charge} is written as
\begin{equation}\label{pc1}n_\ell({\bf r})\,=\,\sum_{j,\Gamma,k}\;c_{\Gamma k}\,Y_{\Gamma,k}({\bf r}-{\bf R}_{\ell j})\end{equation} where $\Gamma$ denotes an Irrep of the Wyckoff symmetry point-group and $k$ indexes its rows; $Y_{\Gamma k}$ is a SA harmonic function. The expansion coefficients $c_{\Gamma k}$ are treated as bona-fide time-dependent dynamical variables, alongside the ionic displacements that they couple to, namely
$$c_{\Gamma k}(t)\,=\,c_{\Gamma k}^{(0)}+\Delta c_{\Gamma k}(t)$$
 The coupling is treated within the adiabatic approximation. 
\begin{figure}[h!]
\begin{center}
\includegraphics[width=0.7\textwidth]{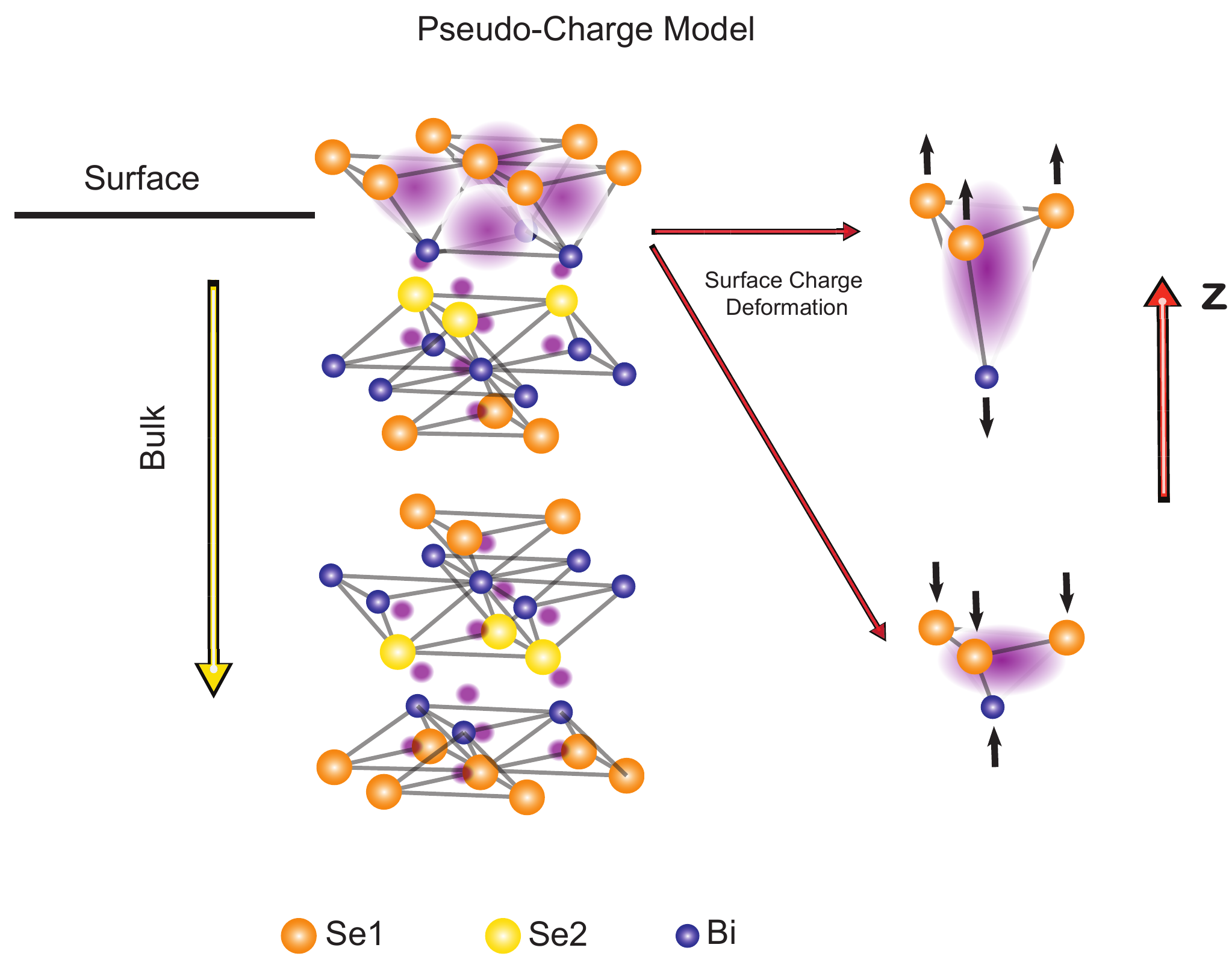}
\caption{\label{psf} {\small A schematic diagram of the lattice structure and the
pseudo-charge model. Two quintuples (Se1-Bi-Se2-Bi-Se1) are shown
in the diagram. The figure indicates that the surface pseudo-charges 
are more spread and easier to deform than their bulk counterparts. The figures to the right, 
show the dipolar pseudo-charge deformations for lattice distortions along the {\bf z} direction.}}
\end{center}
\end{figure}

A Taylor expansion of the total potential energy in terms of ionic displacements $u_\alpha(\ell\kappa)$, where $\alpha$ and $\kappa$ denote Cartesian directions and sublattices, respectively, and pseudo-charge fluctuations $\Delta c_{\Gamma k}$ is given in matrix form by
\begin{equation}\label{pc2}V({\bf r})\,=\,V_0+\frac{1}{2}\,\Bigl[{\bf u}\boldsymbol{\cdot\Phi\cdot}{\bf u}+\left({\bf u}\boldsymbol{\cdot}{\bf T}\boldsymbol{\cdot}\Delta{\bf c}+\text{h.c.}\right)+\Delta{\bf c}\boldsymbol{\cdot}{\bf H}\boldsymbol{\cdot}\Delta{\bf c}\Bigr]\end{equation} where ${\bf u}$ and $\Delta{\bf c}$ are vector matrices representing ionic displacements and charge deformations, respectively. $\boldsymbol{\Phi}$ is the ion-ion force-constant matrix, ${\bf T}$ the electron-ion coupling matrix and ${\bf H}$ is the electron-electron coupling matrix. The elements of these matrices are treated as empirical parameters to be fitted to experiment.

The corresponding equations of motion are 
{\small\begin{align}\label{ps0}M_\kappa\,\ddot{u}_\alpha(\ell\kappa)\,&=\,-\sum_{\ell^\prime\kappa^\prime\beta}\,\Phi_{\alpha\beta}\begin{pmatrix}\ell&\ell^\prime\\\kappa&\kappa^\prime\end{pmatrix}\,u_\beta\left(\ell^\prime\kappa^\prime\right)-\sum_{\ell^\prime,j\atop\Gamma,k}\;T_{\alpha\atop\Gamma k}\begin{pmatrix}\ell&\ell^\prime\\\kappa&j\end{pmatrix}\,\Delta c_{\Gamma k}\left(\ell^\prime j\right)\\\label{ps1}m_\Gamma\,\Delta\ddot{c}_{\Gamma k}(\ell j)\,&=\,-\sum_{\ell^\prime\kappa^\prime\alpha}\;T_{\Gamma k\atop\alpha}\begin{pmatrix}\ell&\ell^\prime\\j&\kappa^\prime\end{pmatrix}\,u_\alpha\left(\ell^\prime\kappa^\prime\right)-H_{\Gamma,k}(\ell j)\,\Delta c_{\Gamma k}(\ell j)\end{align}} where $M_\kappa$ and $m_\Gamma$ are the mass of $\kappa^\text{th}$ ion and the electronic effective mass, respectively.

 Invoking the adiabatic approximation, $m_\Gamma=0$, we obtain
\begin{equation}\Delta{\bf c}\,=\,{\bf H}^{-1}\,{\bf T}^T\,{\bf u}\end{equation} while translation invariance against an arbitrary rigid displacement ${\bf u}_0$ leads to
\begin{equation}\Delta{\bf c}\,=\,{\bf H}^{-1}\,{\bf T}^T\,{\bf u}_0\end{equation} and
\begin{equation}{\bf 0}\,=\,\Bigl[\boldsymbol{\Phi}+ {\bf T}{\bf H}^{-1}{\bf T}^T\Bigr]\,\boldsymbol{\cdot}{\bf u}_0\end{equation} so that the effective ion self force constant matrix is expressed as
{\small\begin{equation}\label{pc3}\boldsymbol{\Phi}\begin{pmatrix}\ell&\ell\\\kappa&\kappa\end{pmatrix}\,=\,-\sum^\prime_{\ell^\prime\kappa^\prime}\,\boldsymbol{\Phi}\begin{pmatrix}\ell&\ell^\prime\\\kappa&\kappa^\prime\end{pmatrix}+\sum_{j,\Gamma k}\;{\bf T}_{\Gamma k}\begin{pmatrix}\ell&\ell^\prime\\\kappa&j\end{pmatrix}\boldsymbol{\cdot}{\bf H}^{-1}\boldsymbol{\cdot}{\bf T}^T_{\Gamma k}\begin{pmatrix}\ell^\prime&\ell\\j&\kappa\end{pmatrix}\end{equation}}

 The Fourier transformed equation of motion reduces to
\begin{align}\label{14}\omega^2({\bf q}){\bf u}({\bf q})\,&=\,\boldsymbol{\cal D}({\bf q})\,{\bf u}({\bf q})\notag\\\boldsymbol{\cal D}({\bf q})\,&=\,
\Bigl(\boldsymbol{\Phi}({\bf q})-{\bf T}({\bf q}){\bf H}^{-1}({\bf q}){\bf T}^\dagger({\bf q})\Bigr)
\end{align} where $\boldsymbol{\cal D}({\bf q})$ is the dynamical matrix.

We found that the $6c$ Wyckoff positions of the ${\bf R}\bar{3}m$ space group were the most appropriate to use as centers of pseudo-charge SA multipole expansion. These Wyckoff positions are identified as having coordinates $(0,0,\pm z)$ and  $C_\text{3v}$ point-group symmetry; they define the axes of the tetrahedral pyramids shown in figure \ref{psf}. Among these points, the center of expansion was chosen to be the pyramid center, again, depicted in figure \ref{psf}. $C_\text{3v}$ has Irreps $A_1\,(\text{with SA dipolar harmonic }z)$ and $E\,(\text{with SA dipolar harmonics }x,y)$. In order to minimize the number of empirical constants employed, we opted to include only the $A_1$ SA fluctuations. The right part of figure  \ref{psf} shows a schematic representation of the pseudo-charge deformation associated with $A_1$ symmetry.
\subsubsection*{A. Bulk phonon dispersion curves}
\begin{figure}[h!]
\begin{center}
\includegraphics[width=0.7\textwidth]{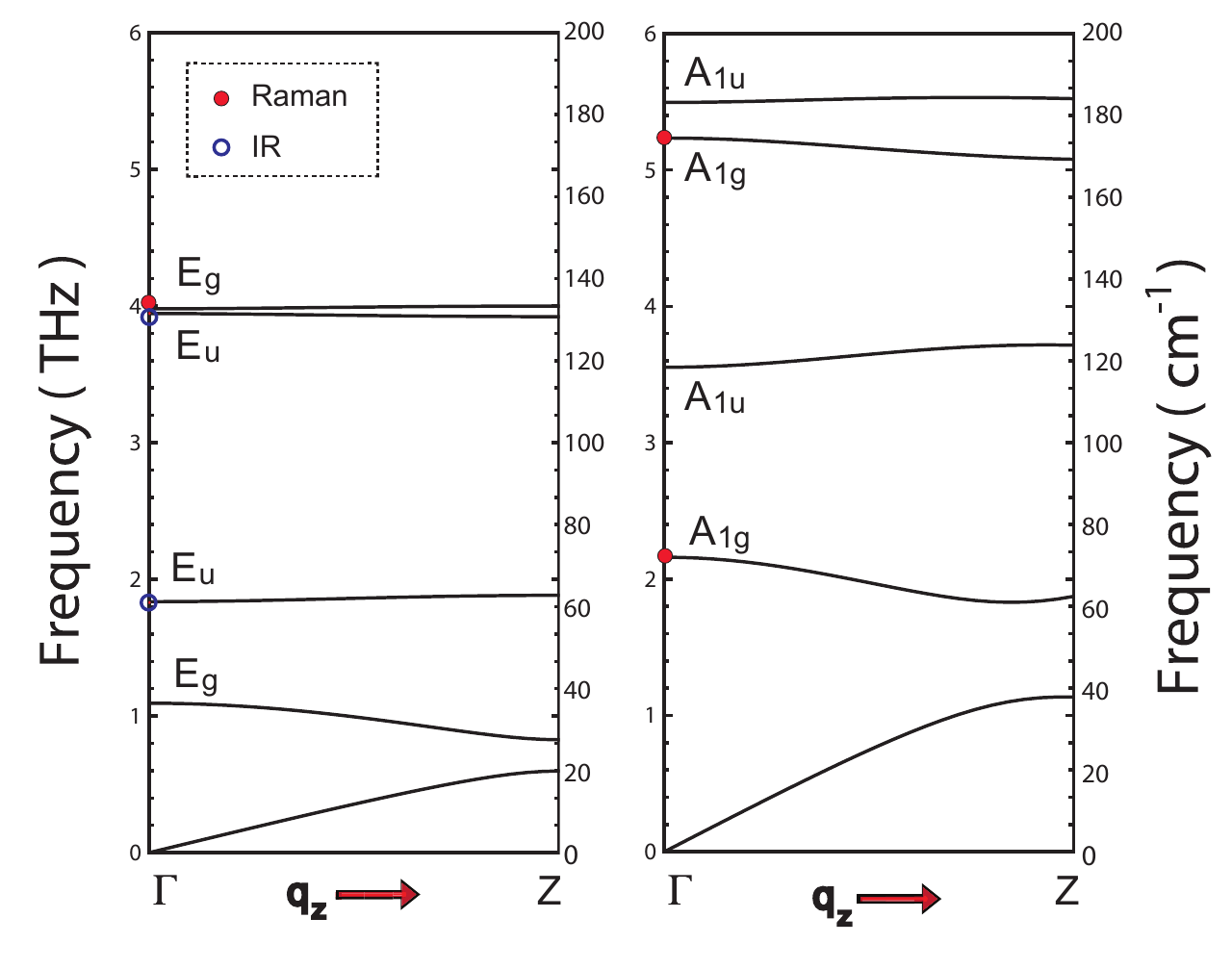}
\caption{\label{bp} {\small Calculated phonon dispersion curves for the bulk of
Bi$_2$Se$_3$ along the $\Gamma$-Z direction, based on the
pseudo-charge model. The experimental data are from
Ref.\cite{Richter}. The red dots correspond to experimental Raman
frequencies, and the blue circles correspond to experimental
infrared (IR) frequencies.}}
\end{center}
\end{figure}
 The initial fitting procedure was carried out for bulk phonons where Raman and IR phonon frequencies are available in the literature \cite{Richter,Landolt}. To our knowledge no neutron scattering data has been reported for Bi$_2$Se$_3$. Central ion-ion interaction potentials, $v(r)$, were adopted; with force-constant matrix elements given by 
$$\Phi_{\alpha\beta}\,=\,A\,\frac{x_\alpha\,x_\beta}{r_0^2}-B\,\left(\frac{x_\alpha\,x_\beta}{r_0^3}-\frac{1}{r_0}\delta_{\alpha\beta}\right)$$
 where 
$$A\,=\,\frac{\partial^2v}{\partial r^2}\Bigr|_{r=r_0},\text{ and } B\,=\,\frac{\partial v}{\partial r}\Bigr|_{r=r_0}$$ where $r_0$ is the equilibrium bond length. A third parameter, $T$, defines the ion-pseudocharge coupling. We used a value of $T_1$ for all pyramids involving Se1-Bi and $T_2$ for those involving Se2-Bi. In the present model, we neglect pseudocharge-pseudocharge interactions, thus rendering the matrix ${\bf H}$ diagonal; its elements are determined from translation invariance constraints.

\begin{table}[h!]
\centering \caption{Best-fit parameters (in N/m) for bulk pseudocharge model, together with experimental (Ref.\cite{Richter,Landolt}) and calculated frequencies.}\vspace{0.1in}\hspace{-0.5in}
\begin{tabular}{|lll|ll|lcl|} \hline \hline\multicolumn{8}{|c|}{}\\
\multicolumn{3}{|c|}{\small Ion-Ion Interaction} &
\multicolumn{2}{|c|}{\small Ion-Pseudocharge Interaction}
&\multicolumn{3}{|c|}{\small Frequencies at $\Gamma$-point (THz)}\\[6pt]  \hline\multicolumn{8}{|c|}{}\\
{\small Bond} &{\small A}    & {\small B}     & {\small Position}              & {\small Constant} &       &{\small Calculated}&{\small Experiment}  \\[6pt]\hline\multicolumn{8}{|c|}{}\\
        &      &       &                       &                      & A$_{1u}$& 5.496 & NA\\
{\small Se1-Se1} & 1.0  & 0.1  & {\small $T_B^1$ (Bi-Se1)}             & 8.07                & A$_{1g}$& 5.235 &  5.235\\
{\small Se1-Bi}  & 13.5 & 1.35 & {\small $T_B^2$ (Bi-Se2)}             & 7.46                & A$_{1u}$& 3.554 & NA\\
{\small Bi-Se2}  & 3.0  & 0.3  &                       &                      & A$_{1g}$& 2.160 &  2.16\\
{\small Bi-Bi}   & 2.0  & 0.2  &                       &                      & E$_{g}$& 3.979  & 3.945\\
        &      &       &                       &                      & E$_{u}$& 3.945  & 3.87-4.02 \\
        &      &       &                       &                      & E$_{u}$& 1.836  & 1.83-1.95 \\
        &      &       &                       &                      & E$_{g}$& 1.094  & NA\\ [4pt] \hline \hline
\end{tabular}
\end{table}

 The best-fit bulk phonon dispersion curves are presented in figure \ref{bp} along the high-symmetry direction $\Gamma$-$Z$. Raman and IR experimental data are depicted as solid red dots and open circles, respectively. The corresponding fitting parameters are given in Table 1, where the ion-ion parameters are specified for each bond-type; the Se1-Se1 bond being the van der Waals inter-quintuple one. We also present both experimental frequencies and their calculated counterparts in Table 1 for comparison. For reference, the bulk BZ is shown in figure \ref{bzs}-a.

\begin{figure}[h!]
\begin{center}
\includegraphics[width=0.8\textwidth]{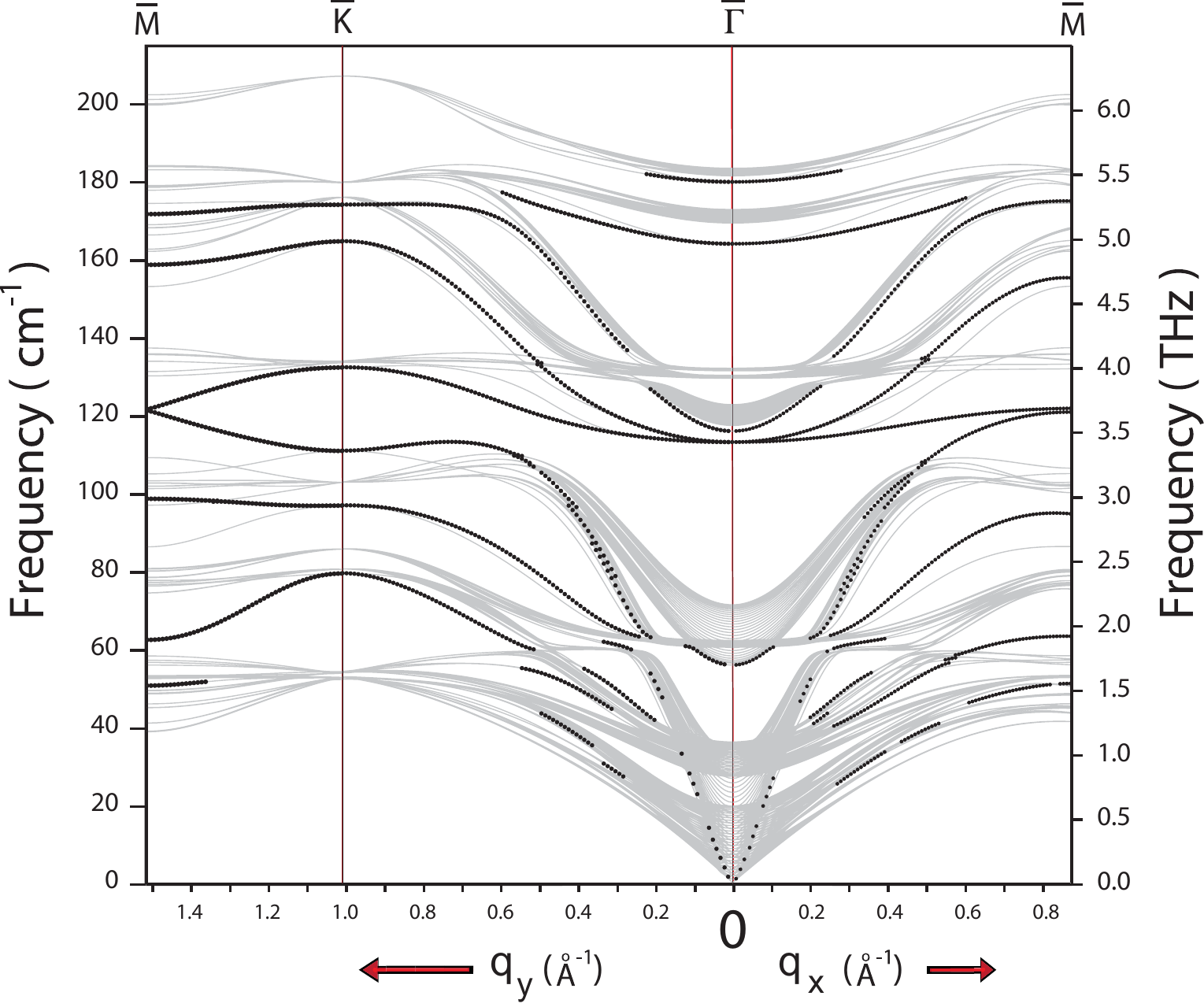}
\caption{\label{spd3} {\small Surface phonon dispersion curves (black dots) along the $\overline{\Gamma}$-$\overline{\text{M}}$ and  $\overline{\Gamma}$-$\overline{\text{K}}$ directions, for\hfill\break $T_S=T_B^1=8.07$ N/m bulk value. The gray background represents the projection of the
bulk bands on the SBZ.}}
\end{center}
\end{figure}

\subsubsection*{B. Surface phonon dispersion curves}
Lattice-dynamics calculations of the surface phonon dispersions based on a pseudocharge lattice-dynamical model \cite{Kaden,Benedek} and a slab geometry containing 30 quintuples were carried out. The (001) Bi$_2$Se$_3$ crystal surface belongs to the $p6mm$ two-dimensional space group. The corresponding surface Brillouin zone (SBZ) is shown in figure \ref{bzs}. The high-symmetry points are $\overline{\Gamma},\ \overline{\text{M}}$ and $\overline{\text{K}}$. The $\overline{\Gamma}$-$\overline{\text{M}}$ direction lies along ${\bf q}_x$, while the $\overline{\Gamma}$-$\overline{\text{K}}$-$\overline{\text{M}}$ line is traced along ${\bf q}_y$. 

To obtain a best-fit to the measured surface phonon dispersion curves the following model changes to bulk parameter values were made: The surface Se-Bi force constant was reduced by 25\% from its bulk value; this is a reasonable change since the non-metallic bonding in the two topmost layers is effectively reduced to allow for the emergence of the metallic electrons. Also a new planar force constant involving surface Se-Se ions was introduced. To account for the metallic deformability of the surface DFQs we reduced the magnitude of the pseudocharge-ion coupling constant in the topmost pyramids involving Se and Bi by about $\frac{\Delta T_S}{T_B^1}=\frac{T_B^1-T_S}{T_B^1}\simeq13\%$. Physically, $\Delta T_S$ accounts for the extra screening provided by the DFQ surface states which is proportional to the corresponding Fermi surface density of states ${\cal D}_F\propto E_F\propto k_F$. Hence $\Delta T_S\propto k_F$. Table \ref{tb2} lists all the modified surface parameters.

However, to underscore the significance of the experimental results and their interpretation in terms of model fitting, we start by presenting the surface phonon dispersions for a model employing the bulk value of the ion-pseudo-charge coupling parameter at the surface,\hfill\break figure \ref{spd3}. The calculated surface phonon dispersion curves are presented in figure \ref{spd2} as black dots, and the gray background represents the projection of the bulk bands onto the SBZ. The important feature to be noted is the presence of a surface Rayleigh branch extending from $\omega=0$ to $\omega\simeq3.7$ THz.

\begin{table}[h!]
\centering \caption{\label{tb2} Best-fit surface parameters (in N/m) of the pseudo-charge slab model calculation.}\vspace{0.1in}
\begin{tabular}{lll|cc} \hline \hline
\multicolumn{3}{p{2.2in}}{Surface Ion-Ion Interaction} &
\multicolumn{2}{p{2.8in}}{Surface Ion-Pseudocharge Interaction} \\
\hline
Ion-Ion &$\qquad$ A    &$\qquad$ B     & Position              & Constant     \\
        &      &       &                       &              \\
Se1-Se1 &$\quad\ $ 2.0  &$\quad\ $ 0.2  & $T_S$(Bi-Se1)             & 7.05        \\
Se1-Bi  &$\quad\ $ 10.0    &$\quad\ $ 1.0   & $\Delta T_S=T_B^1-T_S$=1.02                       &              \\[4pt] \hline \hline
\end{tabular}
\end{table}

\begin{figure}[h!]
\begin{center}
\includegraphics[width=0.8\textwidth]{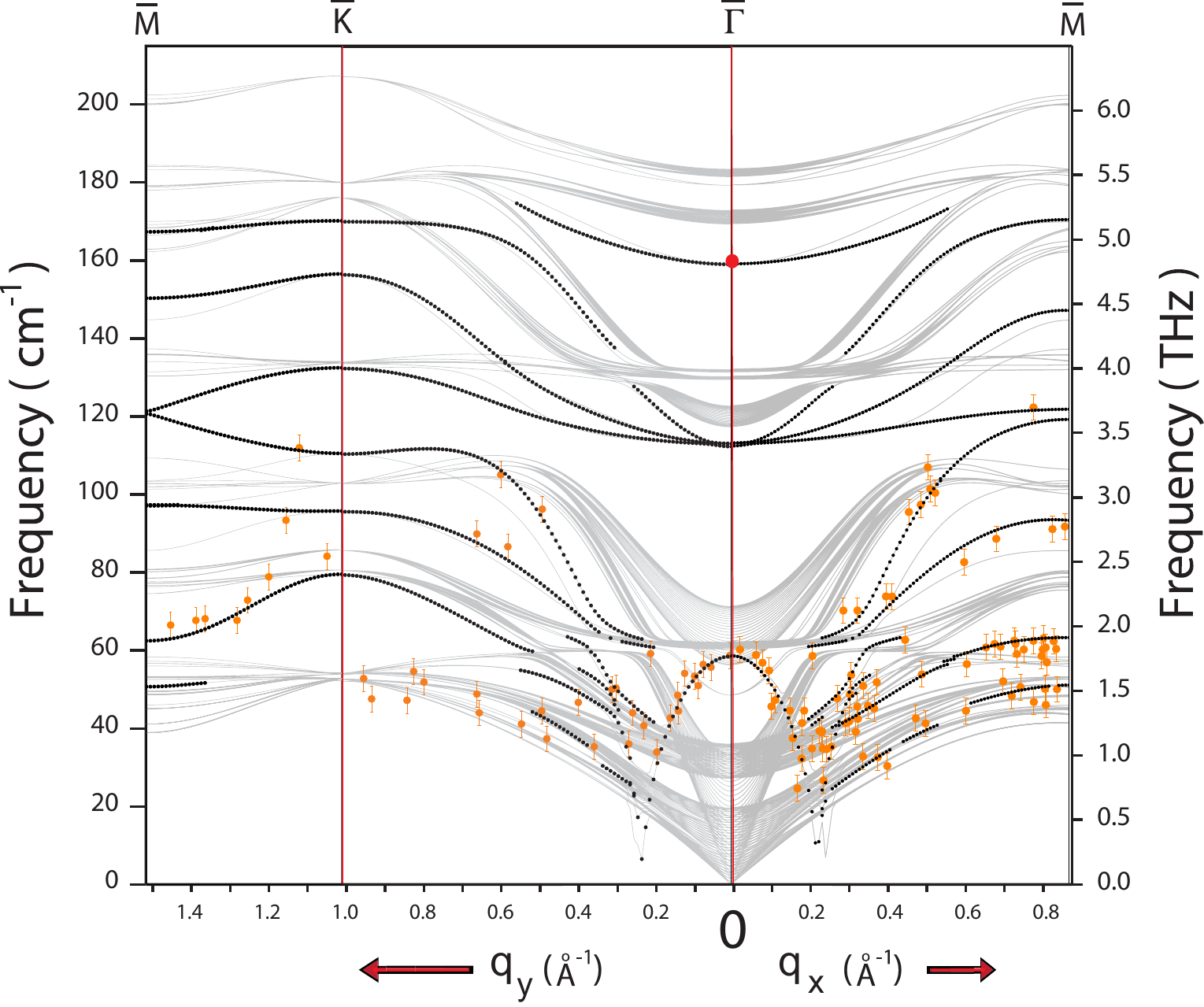}
\caption{\label{spd4} {\small Surface phonon dispersion curves along the $\overline{\Gamma}$-$\overline{\text{M}}$ and  $\overline{\Gamma}$-$\overline{\text{K}}$ directions: Experimental data appear as orange dots with error bars reflecting instrument resolution, while the calculated dispersions curves, using PCM with $T_S = 7.05$ N/m, are represented by black dots. The gray background represents the projection of the bulk bands on the SBZ. The red dot on $\overline{\Gamma}$ at 160 cm$^{-1}$ represents an experimental surface Raman mode reported in Ref.{\cite{Zhao}}. }}
\end{center}
\end{figure}

By contrast, the calculated surface phonon dispersion curves that provide a best-fit for the experimental data, correspond to a reduced value of $T_S$ of 7.05 N/m. The calculated best-fit surface phonon dispersion curves are presented in figure \ref{spd4} as black dots, superimposed on the experimental data which is displayed as solid orange dots with error bars. The single red dot at the $\overline{\Gamma}$-point, at about 5 THz, is a Raman surface frequency reported in \cite{Zhao}.

We focus our attention on two important observations regarding figure \ref{spd}. First, we should discern the absence of a surface Rayleigh branch in both experimental and calculated dispersion curves. Second, we observe the emergence of a ``prominent" dispersion branch centered at the $\overline{\Gamma}$-point with a frequency of 1.8 THz (7.4 meV); the model calculations confirm that it has a  vertical shear (z) polarization at the $\overline{\Gamma}$-point, with $u^z_{Se}/u^z_{Bi}=2.8$. This branch exhibits an anomalous isotropic parabolic dispersion, centered at $\overline{\Gamma}$, and terminates in a V-shaped minimum located at $q\simeq 0.2$ \AA, a value that roughly corresponds to $2k_F$ of the DFQs, and thus signals the manifestation of a strong Kohn anomaly \cite{Kohn}. The isotropy of this branch and its apparent termination at $2k_F$ can be explained by a scenario involving the DFQ surface states, in particular their isotropic Fermi surface. In this scenario, the V-shaped feature marks the boundary between an operative DFQ screening for $q<2k_F$, and its suppression above this value, which is a typical signature of a Kohn anomaly.  Lattice dynamics calculations reveal some bulk penetration of vertical shear modes for $q>2k_F$ reflecting a diminished role of DFQ screening and more compatibility with the insulating bulk. We shall establish below the intimate link between the dispersive character of this branch and the surface DFQ states response to ionic displacements.

\subsection*{
       4. Hamiltonian of the electron-phonon interacting system
        }
\label{sec:Interacting Hamiltonian}
In this section we discuss the renormalization of the surface phonon's frequencies on the $(001)$ surface
of Bi$_2$Se$_3$ due to the electron (DFQ)-phonon interaction in terms of a microscopic model that incorporates the surface Dirac Hamiltonian, the non-interacting surface phonon Hamiltonian and the DFQ-surface phonon interaction Hamiltonian. The latter involves linear coupling of the lattice ionic displacement to the DFQ density. As we shall demonstrate below, the model provides a very good fit to the experimental results for the ``prominent" dispersion branch that has been linked to DFQ screening above. The model also establishes the nature of the screening mechanism to surface phonon excitations, and, thus, complements the analysis provided by the pseudo-charge model.

The system is described by the Hamiltonian
\begin{equation}
\label{eq:total Hamiltonian}
{\cal H} = {\cal H}_{\text{el}} + {\cal H}_{\text{ph}} + {\cal H}_{\text{el-ph}}
\equiv
{\cal H}_{0} + {\cal H}_{\text{el-ph}}\;, 
\end{equation}
whereby ${\cal H}_{0}$ comprises the non-interacting part of the Hamiltonian
while ${\cal H}_{\text{el-ph}}$ describes the interacting part, all of which will
be specified hereafter.

We take the $(001)$ surface of Bi$_2$Se$_3$ to be the $x$-$y$ plane, with area $\mathcal{A}$
normal to the unit vector $\hat{{\mathbf{z}}}$, and consider a
two-dimensional lattice containing $N$ ions, each of mass $M$, which are labeled by the index $j$, and 
assign to each of these atoms an equilibrium position $\mathbf{R}^{(0)}_j$.
A displacement $\mathbf{u}_j$ of the $j$-th ion about its equilibrium position $\mathbf{R}^{(0)}_j$
can be expanded as
\begin{equation}
\mathbf{u}_j
=
\frac{1}{\sqrt{N}}\,\sum_{{\bf q},\gamma}\,\frac{l_{{\bf q},\gamma}}{\sqrt{2}}\,e\,^{\mathrm{i}{\bf q}\cdot{\bf R}^{(0)}_{j}}\,
(b_{{\bf q},\gamma}+b^{\dagger}_{{\bf q},\gamma})\;\hat{\bf e}_{\gamma}(\mathbf q)
\;,	
\end{equation}
where $l_{{\mathbf{q}},\gamma} \equiv \sqrt{\frac{\hbar}{M \omega^{(0)}_{{\bf{q}},\gamma}}}$
and $\hat{\bf e}_{\gamma}(\mathbf q)$ is the polarization vector.
The harmonic modes of vibration are then described by the standard phonon (lattice) Hamiltonian
\begin{equation}
\label{eq:phonon Hamiltonian}
{\cal H}_{\rm{ph}}
=
\sum_{\bf{q},\gamma}\,
\Omega^{(0)}_{\bf{q},\gamma}\,\left( b^{\dagger}_{\bf{q},\gamma}\,b_{\bf{q},\gamma} + \frac{1}{2} \right)\;,	
\end{equation}
where one identifies $b^{\dagger}_{\bf{q},\gamma}$ as the creation operator of a phonon with energy $\Omega^{(0)}_{\bf{q},\gamma} \equiv \hbar\,\omega^{(0)}_{\bf{q},\gamma}$ and polarization index $\gamma$.

The electronic surface states of Bi$_2$Se$_3$ form a two dimensional Dirac metal,
whose low energy physics is well described by the Hamiltonian
\begin{equation}
\label{eq:Dirac Hamiltonian}
{\cal H}_{\rm {el}}
= 
\sum_{\bf{k}}\,
\psi^{\dagger}_{\bf{k}}\,
\left[
\hbar\, \mathrm{v}_{F}\,\hat{\mathbf{z}}\cdot(\mathbf{k}\boldsymbol{\times\sigma})-\mu
\right]
\,
\psi_{\bf{k}}\;,	
\end{equation}
where
$\psi_{\bf{k}} \equiv \displaystyle{\begin{pmatrix}c_{\bf{k}\,\uparrow}\\c_{\,\bf{k}\,\downarrow}\end{pmatrix}}$
is the two-component electron spinor operator at wave vector $\bf{k}$, $\mathrm{v}_{F}$ is the Fermi velocity, $\mu$ is the Fermi energy
(which lies above the Dirac point) and $\boldsymbol{\sigma} = (\sigma_1,\sigma_2)$ is the vector containing the first two Pauli matrices. The Dirac Hamiltonian~(\ref{eq:Dirac Hamiltonian}) is diagonal
in the helicity basis $\Psi_{{\bf k}}\,=\,\displaystyle{\begin{pmatrix} \gamma^{+}_{{\bf k}}\\\gamma^{-}_{{\bf k}}\end{pmatrix}}$:
\begin{equation}
\begin{split}
&
\Psi_{{\bf k}}\,=\,U_{{\bf k}}\,\psi_{{\bf k}}	
\\
&
U_{{\bf k}}=\frac{1}{\sqrt{2}}\,\begin{pmatrix}\text{i}\,e\,^{\text{i}\,\varphi_{{\bf k}}}&1\\-\text{i}\,e\,^{\text{i}\,\varphi_{{\bf k}}}&1\end{pmatrix}
,\quad \varphi_{{\bf k}} \equiv \arctan{\left(\frac{k_{y}}{k_{x}}\right)}\;,
\end{split}
\end{equation}
yielding
\begin{equation}
\label{eq:Dirac Hamiltonian helicity basis}
{\cal H}_{\rm {el}}
= 
\sum_{\bf{k}}\,
\sum_{\alpha = \pm}\,
\xi^{\lambda}_{\bf{k}}\,
(\gamma^{\alpha}_{\bf{k}})^{\dagger}\,
\gamma^{\alpha}_{\bf{k}},
\quad
\xi^{\alpha}_{\bf{k}} = \alpha\,\hbar\mathrm{v}_{F}\,|\bf{k}| - \mu	\;.
\end{equation}
  
We consider an interaction between the electron density and the displacement $\mathbf{u}_{j}$ 
(where $\mathbf{u}_j$ has both in-plane and out-of-plane components)
of the $j^{\rm th}$ ion about its in-plane equilibrium position $\mathbf{R}^{(0)}_{j}$.
The electron-phonon interaction can be generically written as
\begin{equation}
\label{eq:electron-phonon interaction}
{\cal H}_{\rm{el-ph}}
=
\int\,\mathrm{d}^{2}{\bf r}\,\rho_{\rm{el}}({\bf r})\,\sum_{j=1}^{N}\,\boldsymbol{\lambda}(\,{\bf r}-{\bf R}^{(0)}_{j}\,)\,\boldsymbol{\cdot}\mathbf{u}_j\;.
\end{equation}
Here, $\rho_{\rm{el}}({\bf r}) = \displaystyle{\sum_{\sigma = \uparrow,\downarrow}}\,c^{\dagger}_{\sigma}({\bf r})c_{\sigma}({\bf r})$
is the electron surface density operator and $\boldsymbol{\lambda}(\,{\bf r}-{\bf R}^{(0)}_{j}\,)$ 
is a position dependent vector function (with units of energy per length) characterizing the electron-phonon coupling.

We now proceed to represent the interaction Hamiltonian~(\ref{eq:electron-phonon interaction})
in Fourier space. This is achieved by expressing:
\begin{align}
\boldsymbol{\lambda}(\,{\bf r}-{\bf R}^{(0)}_{j}\,)&
=
\frac{1}{\mathcal{A}}\,
\sum_{{\bf q}}\,\boldsymbol{\lambda}_{{\bf q}}\;e\,^{\mathrm{i}{\bf q}\cdot({\bf r}-{\bf R}^{(0)}_{j})}\;\\
\rho_{\rm{el}}({\bf r}) = \sum_{\sigma = \uparrow,\downarrow}\,c^{\dagger}_{\sigma}({\bf r})c_{\sigma}({\bf r})&
=
\frac{1}{\mathcal{A}}\,\sum_{\sigma = \uparrow,\downarrow}\,\sum_{{\bf k},{\bf q}}\,e\,^{-\mathrm{i}{\bf q}\cdot{\bf r}}\,
c^{\dagger}_{{\bf k}+{\bf q},\,\sigma}c_{{\bf k},\,\sigma}\;.
\end{align}
With that, the electron-phonon interaction Hamiltonian~(\ref{eq:electron-phonon interaction}) becomes
\begin{equation}
{\cal H}_{\rm{el-ph}}
=
\frac{1}{\sqrt{\mathcal{A}}}\,\sum_{\sigma = \uparrow,\downarrow}\,\sum_{{\bf k}}\sum_{{\bf q},\gamma}\,{\rm g}_{{\bf q},\gamma}\,
c^{\dagger}_{{\bf k}+{\bf q},\,\sigma}c_{{\bf k},\,\sigma}\;\hat{A}_{{\bf q},\gamma}\;,
\end{equation}
where $\hat{A}_{{\bf q},\gamma}\,\equiv\,(b_{{\bf q},\gamma}+b^{\dagger}_{{\bf q},\gamma})$ and
the electron-phonon coupling constant
\begin{equation}
\label{eq:electron-phonon coupling constant general}
{\rm g}_{{\bf q},\gamma} = \sqrt{\frac{N\hbar}{2 M \mathcal{A}\, \omega^{(0)}_{{\bf q},\gamma}}}\;
\boldsymbol{\lambda}_{\mathbf q} \boldsymbol{\cdot}\hat{\bf e}_{\gamma}(\mathbf q)	
\equiv
 \sqrt{\frac{N\hbar}{2 M \mathcal{A}\, \omega^{(0)}_{{\bf q},\gamma}}}\;
\lambda_{{\mathbf q},\gamma}\;.
\end{equation}

The renormalized phonon frequencies can be found by determining the phonon propagator. We work
in the random phase approximation (RPA), where the phonon propagator is given by
\begin{equation}
\label{eq:RPA phonon propagator}
\mathcal{D}_{\gamma}({\bf q}, \mathrm{i}\omega_{n} )
=
\frac{\mathcal{D}_{\gamma}^{(0)}({\bf q}, \mathrm{i}\omega_{n})}
{
1-
\mathcal{D}_{\gamma}^{(0)}({\bf q}, \mathrm{i}\omega_{n})\,
|\text{g}_{{\bf q}}|^2\,
\frac{\Pi({\bf q}, \mathrm{i}\omega_{n})}{\varepsilon({\bf q}, \mathrm{i}\omega_{n})}
}\;.
\end{equation}
In (\ref{eq:RPA phonon propagator}),
\begin{equation}
\label{eq:non-interacting phonon propagator}
\mathcal{D}_{\gamma}^{(0)}({\bf q}, \mathrm{i}\omega_{n} )
=
\frac{2\left(\hbar\omega^{(0)}_{{\bf q},\gamma}\right)}{(i\omega_n)^{2}-\left(\hbar\omega^{(0)}_{{\bf q},\gamma}\right)^2}\;.
\end{equation} 
is the non-interacting phonon Matsubara Green function, 
\begin{equation}
\label{eq:RPA polarization}
\Pi({\bf q}, \mathrm{i}\omega_{n})
=
\frac{1}{\mathcal{A}}\,\frac{1}{\beta}\,
\sum_{\mathrm{i}\Omega_{n}}\,
\sum_{{\bf p}}\,
\mathrm{Tr}
\left[
G^{(0)}({\bf p}+{\bf q},\mathrm{i}\Omega_{n}+\mathrm{i}\omega_{n})\,
G^{(0)}({\bf p},\mathrm{i}\Omega_{n})
\right]\;	
\end{equation}
is the the RPA electron polarization function and
\begin{equation}
\label{eq:dielectric function}
\varepsilon({\bf q}, \mathrm{i}\omega_{n})
=
1-v_c({\bf q}\,)\,\Pi({\bf q}, \mathrm{i}\omega_{n})\;
\end{equation}
is the RPA dielectric function. $v_c({\bf q}\,) = \frac{e^2}{2\varepsilon_{0}|{\bf q}|}$ 
is the two-dimensional Fourier transform of the electron-electron Coulomb 
interaction potential $v_c({\bf r}) = \frac{e^2}{4\pi\varepsilon_{0}|{\bf r}|}$.

In (\ref{eq:RPA polarization}) the 
non-interacting electron Green's function is defined as
\begin{equation}
G^{(0)}_{\sigma,\sigma^\prime}({\bf p}, \mathrm{i}\omega_{n} )
=
-
\int^{\beta}_{0}\,\mathrm{d}\tau\,
e\,^{\mathrm{i}\omega_{n}\tau}\,
\left<\,T_{\tau}\,
c_{{\bf p},\,\sigma}(\tau)\,
c^{\dagger}_{{\bf p},\,\sigma^\prime}(0)\,
\right>_{0}\;,
\end{equation}
where $\beta \equiv k_{B}\,T$ and $\mathrm{Tr}$ acts on the spin degrees of freedom $\sigma, \sigma^\prime = \uparrow, \downarrow$.

With (\ref{eq:non-interacting phonon propagator}), the RPA phonon propagator (\ref{eq:RPA phonon propagator}) 
can be written as
\begin{equation}
\label{eq:RPA phonon propagator modified}
\mathcal{D}_{\gamma}({\bf q}, \mathrm{i}\omega_{n} )
=
\frac{2(\hbar\omega^{(0)}_{{\bf q},\gamma})}
{
(\mathrm{i}\omega_{n})^2 - (\hbar\omega^{(0)}_{{\bf q},\gamma})^2 
-2(\hbar\omega^{(0)}_{{\bf q},\gamma})|\text{g}_{{\bf q},\gamma}|^2\,
\frac{\Pi({\bf q}, \mathrm{i}\omega_{n})}{\varepsilon({\bf q}, \mathrm{i}\omega_{n})}
}\;.
\end{equation}
The renormalized phonon frequency $\omega_{{\bf q},\gamma}$ is obtained 
by seeking the poles of (\ref{eq:RPA phonon propagator modified}), which then gives
\begin{equation}
(\hbar\omega_{{\bf q},\gamma})^2
=
(\hbar\omega^{(0)}_{{\bf q},\gamma})^2
+2(\hbar\omega^{(0)}_{{\bf q},\gamma})|\text{g}_{{\bf q},\gamma}|^2\,
\frac{\Pi({\bf q}, \omega_{{\bf q},\gamma})}{\varepsilon({\bf q}, \omega_{{\bf q},\gamma})}\;,	
\end{equation}
where the analytic continuation $\mathrm{i}\omega_{n} \rightarrow \omega + \mathrm{i}0^+$ is performed
in order to obtain the retarded response functions.
With (\ref{eq:electron-phonon coupling constant general}),
the RPA phonon propagator can finally be cast in the form
\begin{equation}
\label{eq:self-consistent renormalized frequency}
(\hbar\omega_{{\bf q},\gamma})^2
=
(\hbar\omega^{(0)}_{{\bf q},\gamma})^2 
+\frac{\hbar^2}{M\mathfrak{A}}\,|\lambda_{{\bf q},\gamma}|^{2}\,
\frac{\Pi({\bf q}, \omega_{{\bf q},\gamma})}{\varepsilon({\bf q}, \omega_{{\bf q},\gamma})}\;,
\end{equation}
where $\mathfrak{A}$ is the primitive cell area.

\begin{figure}[h!]
\begin{center}
\includegraphics[width=0.5\textwidth]{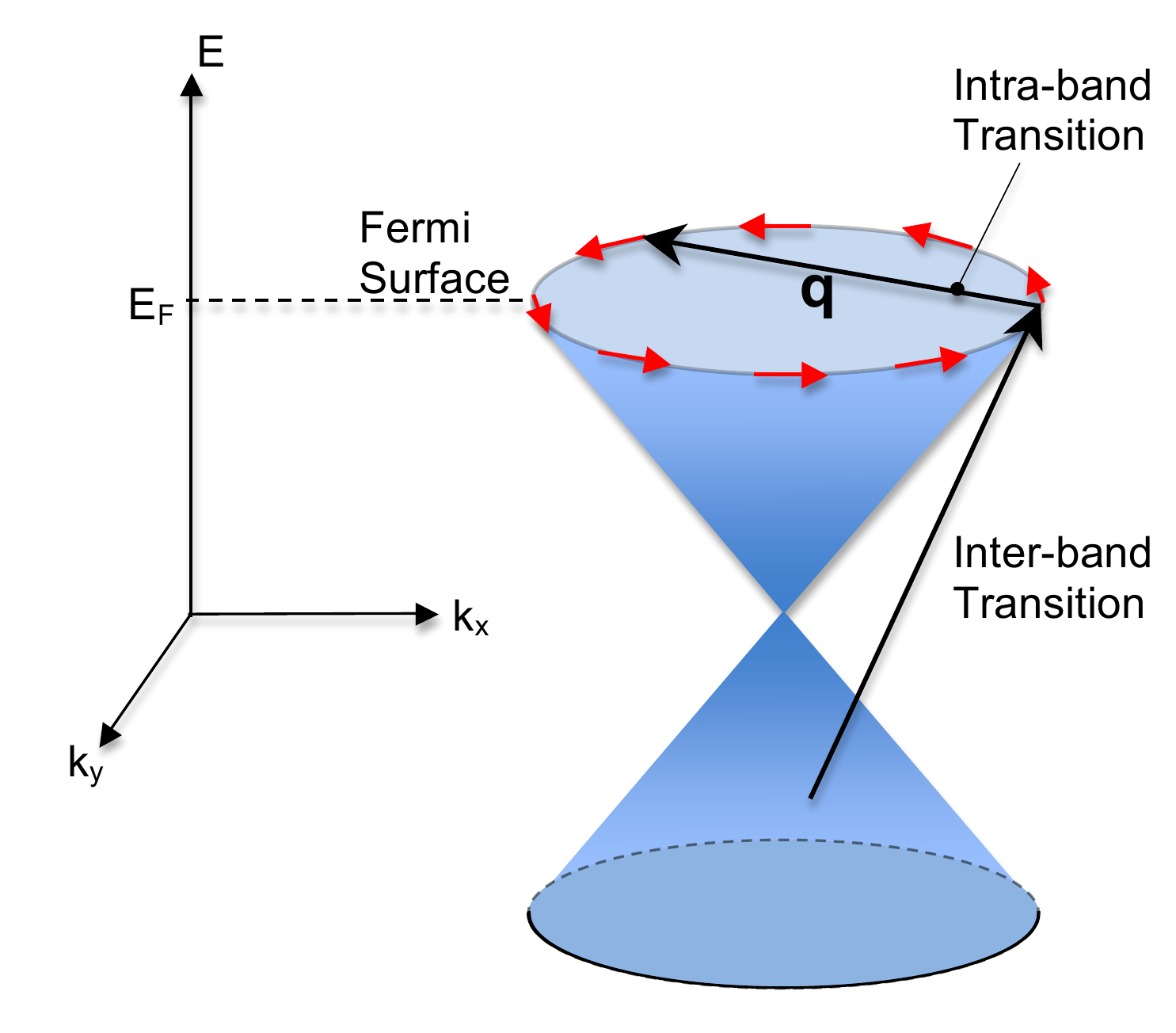}
\caption{\label{DC1} {\small Intra- and Inter-band transitions of DFQs that contribute to the renormalization of the prominent surface phonon branch. ${\bf q}$ is the phonon wave vector.}}
\end{center}
\end{figure}

The polarization function (\ref{eq:RPA polarization}) can be conveniently decomposed into two contributions, one due to the intra-band and the other due to inter-band excitations as shown in figure \ref{DC}:
\begin{equation}
\Pi({\bf q},\mathrm{i}\omega_{n})
=
\Pi_{\,\text{intra}}({\bf q},\mathrm{i}\omega_{n})
+
\Pi_{\,\text{inter}}({\bf q},\mathrm{i}\omega_{n})\;.	
\end{equation}
Upon performing the Matsubara sums in
(\ref{eq:RPA polarization}), we obtain 
\begin{eqnarray}
\Pi_{\,\text{intra}}({\bf q},\mathrm{i}\omega_{n})
=
\int_{|{\bf k}|<k_{\Lambda}} \frac{\mathrm{d}^2{\bf k}}{(2\pi)^{2}}\,
F^{\,\text{intra}}_{{{\bf k},{\bf q}}}\,
\left[
\frac{
n_{F}\left(\xi^{+}_{\mathbf{k}+\mathbf{q}}\right) - n_{F}\left(\xi^{+}_{\mathbf{k}}\right)
}
{\xi^{+}_{\mathbf{k}+\mathbf{q}} - \xi^{+}_{\mathbf{k}} -\mathrm{i}\omega_{n}}+
\frac{
n_{F}\left(\xi^{-}_{\mathbf{k}+\mathbf{q}}\right) - n_{F}\left(\xi^{-}_{\mathbf{k}}\right)
}
{\xi^{-}_{\mathbf{k}+\mathbf{q}} - \xi^{-}_{\mathbf{k}} -\mathrm{i}\omega_{n}}
\right]\;
\end{eqnarray}
and
\begin{eqnarray}
\Pi_{\,\text{inter}}({\bf q},\mathrm{i}\omega_{n})
=
\int_{|{\bf k}|<k_{\Lambda}} \frac{\mathrm{d}^2{\bf k}}{(2\pi)^{2}}\,
F^{\,\text{inter}}_{{{\bf k},{\bf q}}}\,
\left[
\frac
{
n_{F}\left(\xi^{+}_{\mathbf{k}+\mathbf{q}}\right)-n_{F}\left(\xi^{-}_{\mathbf{k}}\right) 
}
{
\xi^{+}_{\mathbf{k}+\mathbf{q}}-\xi^{-}_{\mathbf{k}} -\mathrm{i}\omega_{n}
}
+
\frac
{
n_{F}\left(\xi^{-}_{\mathbf{k}+\mathbf{q}}\right) - n_{F}\left(\xi^{+}_{\mathbf{k}}\right)  
}
{
\xi^{-}_{\mathbf{k}+\mathbf{q}} - \xi^{+}_{\mathbf{k}} - \mathrm{i}\omega_{n}
}
\right]\;.
\end{eqnarray}
$k_{\Lambda}$ denotes a momentum cutoff indicating the penetration of the metallic surface states into the bulk states, 
$n_{F}(\xi) \equiv (1+e^{\beta\,\xi})^{-1}$ is the Fermi-Dirac distribution
and, the chirality factors are
\begin{equation}
\begin{split}
&
F^{\,\text{intra}}_{{{\bf k},{\bf q}}}\,
=
\frac{1}{2}\,
\left(
1+\,\frac{{\bf k}\cdot({\bf k}+{\bf q})}
{|{\bf k}||{\bf k}+{\bf q}|}
\right)	
\\
\\
&
F^{\,\text{inter}}_{{{\bf k},{\bf q}}}\,
=
\frac{1}{2}\,
\left(
1-\,\frac{{\bf k}\cdot({\bf k}+{\bf q})}
{|{\bf k}||{\bf k}+{\bf q}|}
\right)	
\end{split}
\end{equation}
For $T=0$ and $\mu>0$ the polarization expressions take the form
\begin{eqnarray}
\Pi^{\text{intra}}({\bf q},\mathrm{i}\omega_{n})
=
\int_{|{\bf k}|<k_{\Lambda}} \frac{\mathrm{d}^2{\bf k}}{(2\pi)^{2}}\,
F^{\text{intra}}({{\bf k},{\bf q}})\,
\left[
\frac{
\Theta(\mu - v_{F}|{\bf k}+{\bf q}|) - \Theta(\mu - v_{F}|{\bf k}|)
}
{v_{F}(|{\bf k}+{\bf q}|- |{\bf k}|) -\mathrm{i}\omega_{n}}
\right]\;
\end{eqnarray}
and
\begin{eqnarray}
\Pi^{\text{inter}}({\bf q},\mathrm{i}\omega_{n})
=
\int_{|{\bf k}|<k_{\Lambda}} \frac{\mathrm{d}^2{\bf k}}{(2\pi)^{2}}\,
F^{\text{inter}}({{\bf k},{\bf q}})\,
\left[
\frac
{
\Theta(\mu - v_{F}|{\bf k}+{\bf q}|)-1 
}
{
v_{F}(|{\bf k}+{\bf q}| + |{\bf k}|) -\mathrm{i}\omega_{n}
}
+
\frac
{
\Theta(\mu - v_{F}|{\bf k}|)-1 
}
{
v_{F}(|{\bf k}+{\bf q}| + |{\bf k}|) +\mathrm{i}\omega_{n}
}
\right]
\nonumber
\end{eqnarray}

In order to solve (\ref{eq:self-consistent renormalized frequency}) self consistently,
we need to specify the momentum dependence of the coupling function $\boldsymbol{\lambda}_{\mathbf{q},\gamma}$.
We argue that in the range of $\left|{\bf q}\right|\le 2k_F$, the ionic screened potential $V({\bf q})$ has very weak dependence on $q$, since $2k_F < q_{\rm TF}=\frac{e^2\,k_{F}^2}{4\pi\varepsilon_{0}\,E_{F}}=0.5$ \AA$^{-1}$ the Thomas-Fermi wave vector. Moreover, invoking the sagittal-plane symmetry \footnote{The sagittal-plane is defined by the surface wave vector ${\bf q}$ and the surface normal. The only symmetry operation associated with ${\bf q}$ along high-symmetry directions is a reflection through the sagittal-plane, having even and odd parity irreducible representations.} classification of the surface phonon modes, we specify the general polarization vector of the even parity mode as
\begin{equation}
\begin{split}
&
\hat{\bf e}_{\gamma}(\mathbf{q})
=
\hat{\bf e}_{\bot}(\mathbf{q})
+
\hat{\bf e}_{\|}(\mathbf{q})\;,
\\
&
\hat{\bf e}_{\bot}(\mathbf{q}) \sim \hat{\mathbf{z}},
\quad
\hat{\bf e}_{\|}(\mathbf{q}) \sim \hat{\mathbf{q}}\;.
\end{split}
\end{equation}
The component $\hat{\bf e}_{\bot}(\mathbf{q})$ accounts for the vertical sheer component (ionic oscillations
 in a direction perpendicular to the plane), while $\hat{\bf e}_{\|}(\mathbf{q})$ is the longitudinal component. 

The corresponding coupling constant then reads
\begin{align}
\lambda_{\mathbf{q},\gamma}
&=
\boldsymbol{\lambda}_{\mathbf{q}}\boldsymbol{\cdot}\hat{\bf e}_{\gamma}(\mathbf{q}\,)
\,
=\,
\boldsymbol{\lambda}_{\mathbf{q}}\boldsymbol{\cdot}\hat{\bf e}_{\bot}(\mathbf{q}\,)
+
\boldsymbol{\lambda}_{\mathbf{q}}\boldsymbol{\cdot}\hat{\bf e}_{\|}(\mathbf{q}\,)\notag
\\
&
\equiv
\lambda_{\bot}(\mathbf{q}) + \lambda_{\|}(\mathbf{q})\;.
\end{align}
In view of the near constancy of $V({\bf q})$ for $q\le 2k_F$, and the fact that the electron-phonon coupling involves the gradient of the screened potential, we write
\begin{equation}
\lambda_{\mathbf{q}}
=
\lambda^{(0)}_{\bot}
+
\frac{|\mathbf{q}|}{q_{0}}\lambda^{(0)}_{\|}	
\end{equation}
where $q_0$ will be appropriately chosen as $2k_F$. The Dyson equation now becomes
\begin{equation}
(\hbar
\omega_{{\bf q},\gamma})^2=(\hbar\omega_{{\bf q},\gamma}^{(0)})^2+\frac{\hbar^2}{M\mathfrak{A}}\,(\lambda_{\bot})^2\left(1+\frac{|\mathbf{q}|}{k_F}\frac{\lambda_{\|}}{\lambda_{\bot}}\right)\,\frac{\Pi({\bf q},\omega_{{\bf q},\gamma})}{\varepsilon({\bf q},\omega_{{\bf q},\gamma})}
\end{equation}
to first order in $q$.

\begin{figure}[h!]
\begin{center}
\includegraphics[width=0.7\textwidth]{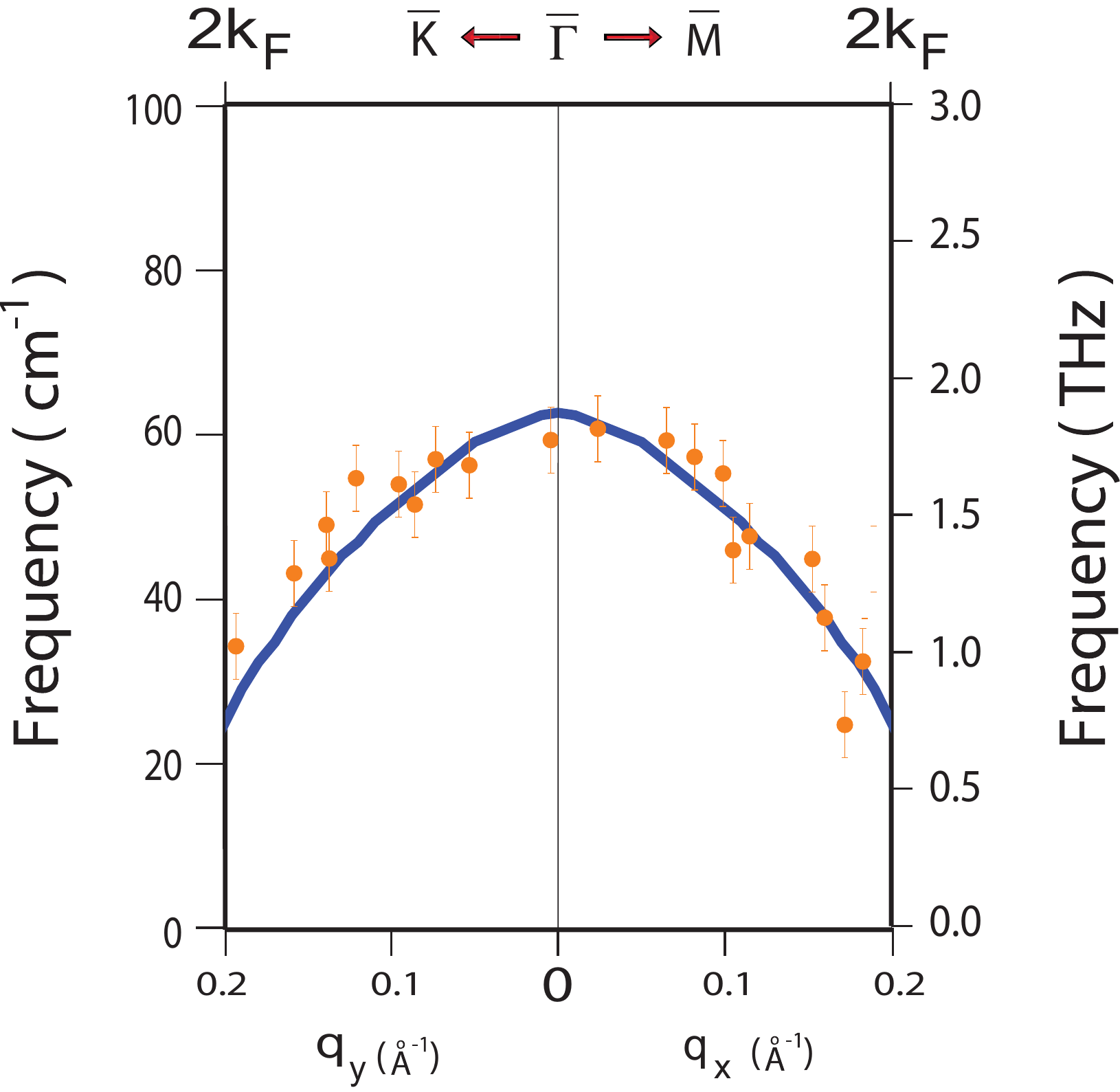}
\caption{\label{Th1} {\small Renormalized surface topological phonon dispersion curve, superimposed on the corresponding experimental data.}}
\end{center}
\end{figure}

The model depends on two parameters: the bare phonon frequency and the Fermi wave vector. We identified the former with the experimental value of $\omega({\bf q}=0)=1.8$ THz, where the DFQ response vanishes. $k_F$ was derived from a sample carrier concentration of $-1.9\times10^{19}/{\rm cm}^3$, obtained from Hall measurements, which correspond to a Fermi energy of about 300 meV and a Fermi wavevector of $k_F=0.1$ \AA$^{-1}$, which is consistent with previous values reported for photoemission measurements. The coupling parameters $\lambda^{(0)}_{\bot}$ and $\lambda^{(0)}_{\|}$ were left as fitting variables. Solutions for $\omega({\bf q})$ were obtained by carrying out the integrals for $\Pi({\bf q})$, and solving Eq.~(\ref{eq:self-consistent renormalized frequency}) iteratively. The calculated best-fit dispersion curve is shown in figure \ref{Th1}, superimposed on the experimental data.

The best fit parameters are
\begin{equation}
\frac{\hbar^2}{M\mathfrak{A}}\,(\lambda_{\bot})^2=10^7 \
\left({\rm meV}\right)^3\cdot{\textup{\AA}}^2, \qquad
\frac{\lambda_{\|}}{\lambda_{\bot}}=0.65
\end{equation}
with $\frac{\hbar^2}{M\mathfrak{A}}=4\times 10^{-3}$ meV, we obtain
$$\lambda_{\bot}=50 \ eV\cdot{\textup{\AA}}$$ this yields a real space value of 3.4 eV/{\textup{\AA}}, which is quite reasonable, since it falls within the range of typical Coulombic interactions.

We should note that recently, Thalmeier \cite{Thalmeier} proposed a model based on acoustic-phonon perturbations, derived from the Dirac fermion Hamiltonian, but these perturbations are too weak to account for the observed dispersion. Moreover, as we demonstrated above, a TI actually presents a composite system consisting of an "ultra-thin metallic film" and an underlying insulating substrate. As such, the phonons in the metallic film can hardly penetrate into the substrate bulk in order to establish an acoustic Rayleigh branch. However, the unique feature of this system is that for $q>2k_F$, as DFQ screening becomes gradually suppressed, the surface film becomes almost homogeneous with the underlying substrate, and the corresponding modes gradually penetrate into the bulk, ushering the V-shaped dispersion. We designate this unique behavior as a {\sl strong} Kohn anomaly. Moreover, we should emphasize here that theoretical modeling and analysis of the interaction between the surface phonons and the DFQs demonstrate that  the linear dispersion and isotropy of the DFQs are responsible for the profile of the prominent surface phonon branch for $q\le 2k_F$.

\bibliographystyle{naturemag}
\bibliography{ref}

\begin{thebibliography}{10}
\expandafter\ifx\csname url\endcsname\relax
  \def\url#1{\texttt{#1}}\fi
\expandafter\ifx\csname urlprefix\endcsname\relax\def\urlprefix{URL }\fi
\providecommand{\bibinfo}[2]{#2}
\providecommand{\eprint}[2][]{\url{#2}}

\bibitem{Hasan}
\bibinfo{author}{Hasan, M.~Z.} \& \bibinfo{author}{Kane, C.~L.}
\newblock \bibinfo{title}{Colloquium: Topological insulators}.
\newblock \emph{\bibinfo{journal}{Rev. Mod. Phys.}}
  \textbf{\bibinfo{volume}{82}}, \bibinfo{pages}{3045--3067}
  (\bibinfo{year}{2010}).

\bibitem{Qi}
\bibinfo{author}{Qi, X.-L.} \& \bibinfo{author}{Zhang, S.-C.}
\newblock \bibinfo{title}{Topological insulators and superconductors}.
\newblock \emph{\bibinfo{journal}{Rev. Mod. Phys.}}  (\bibinfo{year}{2011}).

\bibitem{Moore1}
\bibinfo{author}{Moore, J.~E.}
\newblock \bibinfo{title}{{The birth of topological insulators}}.
\newblock \emph{\bibinfo{journal}{Nature}} \textbf{\bibinfo{volume}{464}},
  \bibinfo{pages}{194--198} (\bibinfo{year}{2010}).

\bibitem{Fu1}
\bibinfo{author}{Fu, L.} \& \bibinfo{author}{Kane, C.~L.}
\newblock \bibinfo{title}{Topological insulators with inversion symmetry}.
\newblock \emph{\bibinfo{journal}{Phys. Rev. B}} \textbf{\bibinfo{volume}{76}},
  \bibinfo{pages}{045302} (\bibinfo{year}{2007}).

\bibitem{Hasan2}
\bibinfo{author}{Hasan, M.~Z.} \& \bibinfo{author}{Moore, J.~E.}
\newblock \bibinfo{title}{Three-dimensional topological insulators}.
\newblock \emph{\bibinfo{journal}{Annual Review of Condensed Matter Physics}}
  \textbf{\bibinfo{volume}{2}}, \bibinfo{pages}{55--78} (\bibinfo{year}{2011}).

\bibitem{Kane}
\bibinfo{author}{Kane, C.~L.} \& \bibinfo{author}{Mele, E.~J.}
\newblock \bibinfo{title}{{\text Z}$_{2}$ topological order and the quantum
  spin hall effect}.
\newblock \emph{\bibinfo{journal}{Phys. Rev. Lett.}}
  \textbf{\bibinfo{volume}{95}}, \bibinfo{pages}{146802}
  (\bibinfo{year}{2005}).

\bibitem{Moore2}
\bibinfo{author}{Moore, J.~E.} \& \bibinfo{author}{Balents, L.}
\newblock \bibinfo{title}{Topological invariants of time-reversal-invariant
  band structures}.
\newblock \emph{\bibinfo{journal}{Phys. Rev. B}} \textbf{\bibinfo{volume}{75}},
  \bibinfo{pages}{121306} (\bibinfo{year}{2007}).

\bibitem{Fu3}
\bibinfo{author}{Fu, L.}, \bibinfo{author}{Kane, C.~L.} \&
  \bibinfo{author}{Mele, E.~J.}
\newblock \bibinfo{title}{Topological insulators in three dimensions}.
\newblock \emph{\bibinfo{journal}{Phys. Rev. Lett.}}
  \textbf{\bibinfo{volume}{98}}, \bibinfo{pages}{106803}
  (\bibinfo{year}{2007}).

\bibitem{Roushan}
\bibinfo{author}{{Roushan}, P.} \emph{et~al.}
\newblock \bibinfo{title}{{Topological surface states protected from
  backscattering by chiral spin texture}}.
\newblock \emph{\bibinfo{journal}{Nature}} \textbf{\bibinfo{volume}{460}},
  \bibinfo{pages}{1106--1109} (\bibinfo{year}{2009}).

\bibitem{Zhang3}
\bibinfo{author}{Zhang, T.} \emph{et~al.}
\newblock \bibinfo{title}{Experimental demonstration of topological surface
  states protected by time-reversal symmetry}.
\newblock \emph{\bibinfo{journal}{Phys. Rev. Lett.}}
  \textbf{\bibinfo{volume}{103}}, \bibinfo{pages}{266803}
  (\bibinfo{year}{2009}).

\bibitem{Chis}
\bibinfo{author}{Chis, V.} \emph{et~al.}
\newblock \bibinfo{title}{Large surface charge density oscillations induced by
  subsurface phonon resonances}.
\newblock \emph{\bibinfo{journal}{Phys. Rev. Lett.}}
  \textbf{\bibinfo{volume}{101}}, \bibinfo{pages}{206102}
  (\bibinfo{year}{2008}).

\bibitem{Benedek}
\bibinfo{author}{Benedek, G.} \emph{et~al.}
\newblock \bibinfo{title}{Theory of surface phonons at metal surfaces: recent
  advances}.
\newblock \emph{\bibinfo{journal}{Journal of Physics: Condensed Matter}}
  \textbf{\bibinfo{volume}{22}}, \bibinfo{pages}{084020}
  (\bibinfo{year}{2010}).

\bibitem{Chen}
\bibinfo{author}{Chen, Y.~L.} \emph{et~al.}
\newblock \bibinfo{title}{Experimental realization of a three-dimensional
  topological insulator, $\text{Bi}_2\text{Te}_3$}.
\newblock \emph{\bibinfo{journal}{Science}} \textbf{\bibinfo{volume}{325}},
  \bibinfo{pages}{178--181} (\bibinfo{year}{2009}).

\bibitem{Noh}
\bibinfo{author}{{Noh, H.-J.}} \emph{et~al.}
\newblock \bibinfo{title}{Spin-orbit interaction effect in the electronic
  structure of $\text{Bi}_{2}\text{Te}_{3}$ observed by angle-resolved
  photoemission spectroscopy}.
\newblock \emph{\bibinfo{journal}{EPL}} \textbf{\bibinfo{volume}{81}},
  \bibinfo{pages}{57006} (\bibinfo{year}{2008}).

\bibitem{Zhang1}
\bibinfo{author}{Zhang, W.}, \bibinfo{author}{Yu, R.}, \bibinfo{author}{Zhang,
  H.-J.}, \bibinfo{author}{Dai, X.} \& \bibinfo{author}{Fang, Z.}
\newblock \bibinfo{title}{First-principles studies of the three-dimensional
  strong topological insulators $\text{Bi}_2\text{Te}_3$,
  $\text{Bi}_2\text{Se}_3$ and $\text{Sb}_2\text{Te}_3$}.
\newblock \emph{\bibinfo{journal}{New Journal of Physics}}
  \textbf{\bibinfo{volume}{12}}, \bibinfo{pages}{065013}
  (\bibinfo{year}{2010}).

\bibitem{Zhang2}
\bibinfo{author}{Zhang, H.} \emph{et~al.}
\newblock \bibinfo{title}{{Topological insulators in $\text{Bi}_2\text{Se}_3$,
  $\text{Bi}_2\text{Te}_3$ and $\text{Sb}_2\text{Te}_3$ with a single Dirac
  cone on the surface}}.
\newblock \emph{\bibinfo{journal}{Nature Physics}}
  \textbf{\bibinfo{volume}{5}}, \bibinfo{pages}{438--442}
  (\bibinfo{year}{2009}).

\bibitem{Hsieh3}
\bibinfo{author}{Hsieh, D.} \emph{et~al.}
\newblock \bibinfo{title}{Observation of time-reversal-protected
  single-dirac-cone topological-insulator states in $\text{Bi}_2\text{Te}_3$
  and $\text{Sb}_2\text{Te}_3$}.
\newblock \emph{\bibinfo{journal}{Phys. Rev. Lett.}}
  \textbf{\bibinfo{volume}{103}}, \bibinfo{pages}{146401}
  (\bibinfo{year}{2009}).

\bibitem{Park2}
\bibinfo{author}{Park, K.}, \bibinfo{author}{Heremans, J.~J.},
  \bibinfo{author}{Scarola, V.~W.} \& \bibinfo{author}{Minic, D.}
\newblock \bibinfo{title}{Robustness of topologically protected surface states
  in layering of $\text{Bi}_2\text{Te}_3$ thin films}.
\newblock \emph{\bibinfo{journal}{Phys. Rev. Lett.}}
  \textbf{\bibinfo{volume}{105}}, \bibinfo{pages}{186801}
  (\bibinfo{year}{2010}).

\bibitem{Xia}
\bibinfo{author}{{Xia}, Y.} \emph{et~al.}
\newblock \bibinfo{title}{{Observation of a large-gap topological-insulator
  class with a single Dirac cone on the surface}}.
\newblock \emph{\bibinfo{journal}{Nature Physics}}
  \textbf{\bibinfo{volume}{5}}, \bibinfo{pages}{398--402}
  (\bibinfo{year}{2009}).

\bibitem{Analytis}
\bibinfo{author}{Analytis, J.~G.} \emph{et~al.}
\newblock \bibinfo{title}{Bulk fermi surface coexistence with dirac surface
  state in $\text{Bi}_{2}\text{Se}_{3}$: A comparison of photoemission and
  $\text{S}$hubnikov--de $\text{H}$aas measurements}.
\newblock \emph{\bibinfo{journal}{Phys. Rev. B}} \textbf{\bibinfo{volume}{81}},
  \bibinfo{pages}{205407} (\bibinfo{year}{2010}).

\bibitem{Park}
\bibinfo{author}{Park, S.~R.} \emph{et~al.}
\newblock \bibinfo{title}{Quasiparticle scattering and the protected nature of
  the topological states in a parent topological insulator
  $\text{Bi}_2\text{Se}_3$}.
\newblock \emph{\bibinfo{journal}{Phys. Rev. B}} \textbf{\bibinfo{volume}{81}},
  \bibinfo{pages}{041405} (\bibinfo{year}{2010}).

\bibitem{Kuroda}
\bibinfo{author}{Kuroda, K.} \emph{et~al.}
\newblock \bibinfo{title}{Hexagonally deformed fermi surface of the $3\text{D}$
  topological insulator $\text{Bi}_2\text{Se}_3$}.
\newblock \emph{\bibinfo{journal}{Phys. Rev. Lett.}}
  \textbf{\bibinfo{volume}{105}}, \bibinfo{pages}{076802}
  (\bibinfo{year}{2010}).

\bibitem{Yazyev}
\bibinfo{author}{Yazyev, O.~V.}, \bibinfo{author}{Moore, J.~E.} \&
  \bibinfo{author}{Louie, S.~G.}
\newblock \bibinfo{title}{Spin polarization and transport of surface states in
  the topological insulators $\text{Bi}_2\text{Se}_3$ and
  $\text{Bi}_2\text{Te}_3$ from first principles}.
\newblock \emph{\bibinfo{journal}{Phys. Rev. Lett.}}
  \textbf{\bibinfo{volume}{105}}, \bibinfo{pages}{266806}
  (\bibinfo{year}{2010}).

\bibitem{Support}
\bibinfo{title}{Supplementary material} .

\bibitem{Jayanthi}
\bibinfo{author}{Jayanthi, C.~S.}, \bibinfo{author}{Bilz, H.},
  \bibinfo{author}{Kress, W.} \& \bibinfo{author}{Benedek, G.}
\newblock \bibinfo{title}{Nature of surface-phonon anomalies in noble metals}.
\newblock \emph{\bibinfo{journal}{Phys. Rev. Lett.}}
  \textbf{\bibinfo{volume}{59}}, \bibinfo{pages}{795--798}
  (\bibinfo{year}{1987}).

\bibitem{Kaden}
\bibinfo{author}{Kaden, C.}, \bibinfo{author}{Ruggerone, P.},
  \bibinfo{author}{Toennies, J.~P.}, \bibinfo{author}{Zhang, G.} \&
  \bibinfo{author}{Benedek, G.}
\newblock \bibinfo{title}{Electronic pseudocharge model for the
  $\text{Cu}$(111) longitudinal-surface-phonon anomaly observed by helium-atom
  scattering}.
\newblock \emph{\bibinfo{journal}{Phys. Rev. B}} \textbf{\bibinfo{volume}{46}},
  \bibinfo{pages}{13509--13525} (\bibinfo{year}{1992}).

\bibitem{Richter}
\bibinfo{author}{Richter, W.} \& \bibinfo{author}{Becker, C.~R.}
\newblock \bibinfo{title}{A raman and far-infrared investigation of phonons in
  the rhombohedral $\text{V}_2$-$\text{VI}_3$ compounds
  $\text{Bi}_2\text{Te}_3$, $\text{Bi}_2\text{Se}_3$, $\text{Sb}_2\text{Te}_3$
  and $\text{Bi}_2(\text{Te}_{1-x}\text{Se}_x)_3$ ($\small{0 < x < 1}$),
  $(\text{Bi}_{1-y}\text{Sb}_y)_2\text{Te}_3$ ($\small{0 < y < 1}$)}.
\newblock \emph{\bibinfo{journal}{Physica Status Solidi (b)}}
  \textbf{\bibinfo{volume}{84}}, \bibinfo{pages}{619--628}
  (\bibinfo{year}{1977}).

\bibitem{Landolt}
\bibinfo{title}{Bismuth selenide ($\text{Bi}_2\text{Se}_3$) phonon dispersion,
  phonon frequencies}.
\newblock In \bibinfo{editor}{O.~Madelung, M.~S., U.~R\"ossler} (ed.)
  \emph{\bibinfo{booktitle}{Landolt-B\"ornstein database - Group III Condensed
  Matter}}, vol. \bibinfo{volume}{41C: Non-Tetrahedrally Bonded Elements and
  Binary Compounds I} (\bibinfo{publisher}{Springer}, \bibinfo{year}{2011}).

\bibitem{Zhao}
\bibinfo{author}{Zhao, S. Y.~F.} \emph{et~al.}
\newblock \bibinfo{title}{Fabrication and characterization of topological
  insulator $\text{Bi}_2\text{Se}_3$ nanocrystals}.
\newblock \emph{\bibinfo{journal}{Applied Physics Letters}}
  \textbf{\bibinfo{volume}{98}}, \bibinfo{pages}{141911}
  (\bibinfo{year}{2011}).

\bibitem{Kohn}
\bibinfo{author}{Kohn, W.}
\newblock \bibinfo{title}{Image of the fermi surface in the vibration spectrum
  of a metal}.
\newblock \emph{\bibinfo{journal}{Phys. Rev. Lett.}}
  \textbf{\bibinfo{volume}{2}}, \bibinfo{pages}{393--394}
  (\bibinfo{year}{1959}).

\bibitem{Farzaneh}
\bibinfo{author}{Farzaneh, M.}, \bibinfo{author}{Liu, X.-F.},
  \bibinfo{author}{El-Batanouny, M.} \& \bibinfo{author}{Chou, F.~C.}
\newblock \bibinfo{title}{Structure and lattice dynamics of
  $sr_{2}cuo_{2}cl_{2}(001)$ studied by helium-atom scattering}.
\newblock \emph{\bibinfo{journal}{Phys. Rev. B}} \textbf{\bibinfo{volume}{72}},
  \bibinfo{pages}{085409} (\bibinfo{year}{2005}).

\bibitem{Martini}
\bibinfo{author}{Martini, K.~M.}, \bibinfo{author}{Franzen, W.} \&
  \bibinfo{author}{El-Batanouny, M.}
\newblock \bibinfo{title}{Compact he-atom detector for high-resolution surface
  phonon measurements}.
\newblock \emph{\bibinfo{journal}{Review of Scientific Instruments}}
  \textbf{\bibinfo{volume}{58}}, \bibinfo{pages}{1027--1037}
  (\bibinfo{year}{1987}).

\bibitem{Thalmeier}
\bibinfo{author}{Thalmeier, P.}
\newblock \bibinfo{title}{Surface phonon propagation in topological
  insulators}.
\newblock \emph{\bibinfo{journal}{Phys. Rev. B}} \textbf{\bibinfo{volume}{83}},
  \bibinfo{pages}{125314} (\bibinfo{year}{2011}).

\end{thebibliography}
\subsection*{ Acknowlegments}
 This work is supported by the U.S. Department of Energy under Grants No. DE-FG02-85ER45222 (MEB) and DEFG02-06ER46316 (CC). FCC acknowledges the support from the
National Science Council of Taiwan under project No. NSC 99-2119-M-002-011-MY.
\subsection*{Author Contributions}
X.Z. carried out the experimental HASS measurements with the assistance of S.K. and C.H. X.Z. also wrote the program and carried out the PCM calculations. R.S. and F.C.C. carried out the crystal growing and provided the samples. L.S. and C.C. carried out the theoretical analysis. M.E.B. conceived the idea for searching for a dynamical signature for Dric fermions in the Bi2X3 topological class and was responsible for overall project direction, planning and management.\vspace{0.3in}
\end{document}